%% file: DRAFT_ME_V3.tex
\documentclass[aps,prx,twocolumn,byrevtex,showpacs,floatfix,superscriptaddress]{revtex4-1}

\usepackage{graphicx}
\usepackage{dcolumn}
\usepackage{amsmath}
\usepackage{epsfig}
\usepackage{bm}
\usepackage{amssymb}
\usepackage{hyperref}
\hypersetup{colorlinks=true,linkcolor=blue,citecolor=blue,urlcolor=blue}
\usepackage{microtype}
\allowdisplaybreaks

\begin{document}

\title{Experimental minimum-error quantum-state discrimination in high dimensions}

\author{M. A. Sol\'is-Prosser}
\email{msolisp@udec.cl} 
\affiliation{Center for Optics and Photonics and MSI-Nucleus on Advanced Optics,
Universidad de Concepci\'on, Casilla 4016, Concepci\'on, Chile}
\affiliation{Departamento de F\'isica, Universidad de Concepci\'on, Casilla 160-C, Concepci\'on, Chile}

\author{M. F. Fernandes}
\affiliation{Departamento de F\'isica, Universidade Federal de Minas Gerais, Belo Horizonte, MG 31270-901, Brazil}

\author{O. Jim\'enez}
\affiliation{\text{Departamento de F\'isica, Facultad de Ciencias B\'asicas, Universidad de Antofagasta, Casilla 170, Antofagasta, Chile}}

\author{A. Delgado}
\affiliation{Center for Optics and Photonics and MSI-Nucleus on Advanced Optics,
Universidad de Concepci\'on, Casilla 4016, Concepci\'on, Chile}
\affiliation{Departamento de F\'isica, Universidad de Concepci\'on, Casilla 160-C, Concepci\'on, Chile}

\author{L. Neves}
\email{lneves@fisica.ufmg.br} 
\affiliation{Departamento de F\'isica, Universidade Federal de Minas Gerais, Belo Horizonte, MG 31270-901, Brazil}

\date{\today}

\begin{abstract}  
Quantum mechanics forbids perfect discrimination among nonorthogonal states through a single shot measurement. To optimize this task, many  strategies were devised that later became fundamental tools for quantum information processing. Here, we address the pioneering minimum-error (ME) measurement and give the first experimental demonstration of its application for discriminating nonorthogonal states in high dimensions.  Our  scheme is designed to distinguish symmetric pure states encoded in the transverse spatial modes of an optical field; the optimal measurement is performed by a projection onto the Fourier transform basis of these modes. For dimensions ranging from $D=2$ to $D=21$ and nearly 14000 states tested, the deviations of the experimental results from the theoretical values range from 0.3\% to 3.6\% (getting below 2\% for the vast majority), thus showing the excellent performance of our scheme. This ME measurement is a building block for high-dimensional implementations of many quantum communication protocols, including probabilistic state discrimination,  dense coding with nonmaximal entanglement, and cryptographic schemes. 
\end{abstract}

\maketitle

Quantum mechanics establishes fundamental bounds to our capability of distinguishing among states with nonvanishing overlap: if one is given at random one of two or more nonorthogonal states and asked to identify it from a single shot measurement, it will be impossible to accomplish the task deterministically and with full confidence. This constraint has deep implications both foundational, underlying the debate about the epistemic and ontic nature of quantum states \cite{Spekkens07,Leifer14,Branciard14,Ringbauer15}, and practical, warranting secrecy in quantum key distribution~\cite{Gisin02,Grosshans03}. Beyond that, the problem of discriminating nonorthogonal quantum states plays an important role in quantum information and quantum communications \cite{Reviews}.

A wide variety of measurement strategies have been devised in order to optimize the state discrimination process according to a predefined figure of merit \cite{Reviews}. The pioneering one was the minimum-error (ME) measurement \cite{Holevo73,Yuen75,HelstromBook} where each outcome identifies one of the possible states and the overall error probability is minimized. Other fundamental strategies conceived later \cite{Ivanovic87,Chefles98,Croke06,Jimenez11,Sugimoto12,Herzog12,Bagan12} employ  the ME discrimination in the step next to a transformation taking the input states to more distinguishable ones \cite{Nakahira12}, which enables us to identify them with any desired confidence level (within the allowed bounds) and a maximum success probability. Nowadays, the ME measurement is central to a range of applications, including quantum imaging~\cite{Tan08}, quantum reading~\cite{Pirandola11}, image discrimination~\cite{Nair11}, error correcting codes~\cite{Lloyd11}, and quantum repeaters~\cite{Loock06}, thus stressing its importance.

Closed-form solutions for ME measurements are known only for a few sets of states. One of these is the set of symmetric pure states (defined below) prepared with equal prior probabilities~\cite{Ban97}. Discriminating among them with minimum error sets the bounds on the eavesdropping in some quantum cryptographic schemes \cite{Phoenix00} and is crucial for optimal deterministic and probabilistic realizations of protocols like quantum teleportation~\cite{Neves12}, entanglement swapping~\cite{Prosser14}, and dense coding~\cite{Pati05}, when the quantum channels are nonmaximally entangled. 

Experimental demonstrations of ME discrimination have been provided in continuous variables for two coherent states~\cite{Cook07}, and in two-dimensional Hilbert spaces for sets of two \cite{Barnett97}, three, and four states \cite{Clarke01} encoded in the light polarization, and two states encoded in the $^{14}$N nuclear spin \cite{Waldherr12}. It is of key importance to extend this to high-dimensional quantum systems (qudits) due to the many advantages they offer over qubit-based applications. Qudits provide an increase in the channel capacity for quantum communication \cite{Fujiwara03}, and a higher error rate tolerance and improved security in quantum key distribution \cite{Cerf02,Huber13,Sheridan10}. Moreover, they are a necessary resource for fundamental tests of quantum mechanics, like  contextuality tests \cite{Kochen67}, and for lowering the detection efficiencies required for nonlocality tests \cite{Vertesi10}. Therefore, the ability to perform ME measurements for qudit states will enhance their potential of use in many of these practical applications. It will also be essential for exploring novel protocols of quantum-state discrimination \cite{Jimenez11,Sugimoto12,Herzog12}, bringing positive impacts for quantum communications.

In this Letter, we report the first experimental demonstration of ME discrimination among nonorthogonal states of a single qudit. This is done for equally likely symmetric states in dimensions $D$ ranging from $D=2$ to $D=21$. Using states encoded in $D$ transverse spatial modes of an optical field, known as spatial qudits \cite{Neves05}, we  carried out the experiment for every dimension in that range in a total of 1851 sets of states. Up to small deviations caused by unavoidable experimental imperfections, our  scheme is shown to be optimal, achieving the minimum error probability allowed by quantum mechanics. 

To outline the problem, consider a set of $N$ quantum states $\{|\psi_j\rangle\}_{j=0}^{N-1}$ spanning a $D$-dimensional Hilbert space $\mathcal{H}_D$, with $D\leq N$, defined by
\begin{equation}    \label{eq:sym-states}
|\psi_j\rangle=\sum_{n=0}^{D-1}c_n\omega^{jn}|n\rangle, 
\end{equation}
where $\omega=\exp(2\pi i/N)$, the $c_n$'s are real and nonnegative ($\sum_nc_n^2=1$), and $\{|n\rangle\}_{n=0}^{D-1}$ is an orthonormal basis in $\mathcal{H}_D$. They are symmetric under $\hat{U}=\sum_{l=0}^{D-1}\omega^{l}|l\rangle\langle l|$ since $|\psi_j\rangle=\hat{U}|\psi_{j-1}\rangle=\hat{U}^j|\psi_0\rangle$ and $|\psi_0\rangle=\hat{U}|\psi_{N-1}\rangle$. If each of these states is prepared  with the same {\it a priori} probability $1/N$, they can be identified with minimum error through the measurement \cite{Ban97}
\begin{equation}    \label{eq:Fourier}
\hat{\Pi}_k^{\rm ME}=\hat{\mathcal{F}}_N|k\rangle\langle k|\hat{\mathcal{F}}_N^{-1}\equiv|\mu_k\rangle\langle\mu_k|, 
\end{equation}
where $k\!=0,\ldots,N-1$, and \mbox{$\hat{\mathcal{F}}_N\!=\!\frac{1}{\sqrt{N}}\sum_{m,n=0}^{N-1}\omega^{mn}|m\rangle\langle n|$} is the discrete Fourier transform acting on an $N$-dimensional Hilbert space $\mathcal{H}_N$. For $N=D$ (linearly independent states), this is a projective measurement on $\mathcal{H}_D$. For $N>D$ (linearly dependent states), this is a projective measurement on $\mathcal{H}_N$ that realizes the positive operator valued measure (POVM) for ME discrimination on $\mathcal{H}_D$. This procedure is based on Neumark's theorem \cite{HelstromBook} for implementing POVMs by embedding the system space $\mathcal{H}_D$ into a larger Hilbert space $\mathcal{H}_N$ given by the direct sum $\mathcal{H}_D\oplus\mathcal{H}_{N-D}$, where $\mathcal{H}_{N-D}$ represents the $N-D$ unused extra dimensions of the original system \cite{Chen07}. From Eqs.~(\ref{eq:sym-states}) and (\ref{eq:Fourier}), the probability of obtaining an outcome $\delta_k$ (associated with $\hat{\Pi}_k^{\rm ME}$) if the prepared state was $|\psi_j\rangle$ is given by $P(\delta_k|\psi_j)=|\langle\mu_k|\psi_j\rangle|^2$. Thus, the maximum overall probability of correctly identifying the state will be 
\begin{equation}     \label{eq:Pcorr}
P_{\rm corr} = \frac{1}{N}\sum_{j=0}^{N-1}P(\delta_j|\psi_j) = \frac{1}{N}\left(\sum_{j=0}^{D-1}c_j\right)^2.
\end{equation}
Equivalently, $P_{\rm err}=1-P_{\rm corr}$ will be the ME probability. 

\begin{figure}[tbp]
\centerline{\includegraphics[width=.487\textwidth]{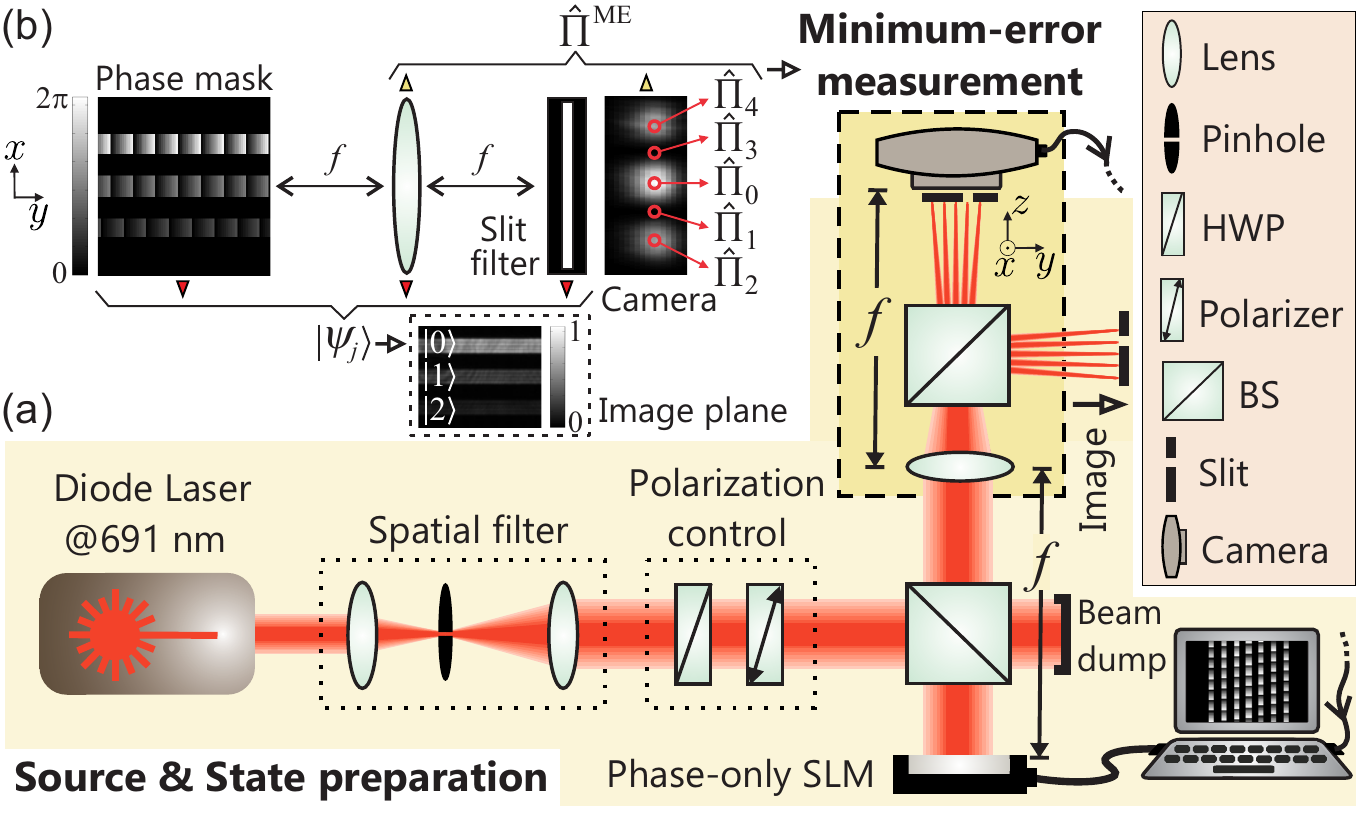}}
\caption{\label{fig:setup}  (a) Experimental setup.  Our light source is a single-mode diode laser operating at 691~nm. The beam profile was spatially filtered, expanded, and collimated so that it was approximately a plane wave with constant phase across the screen of the spatial light modulator (SLM). A half-wave plate (HWP) followed by a polarizer provided a clean vertical polarization (the working direction of the SLM) and acted as a variable attenuator to avoid saturation of the detectors. The first 50:50 beam splitter (BS) enabled the normal incidence of the beam onto the reflective phase-only SLM (Holoeye PLUTO). The modulated beam was transmitted through a converging lens ($f=30$~cm) and  split in two arms by the second BS. At each arm a slit diaphragm was placed nearly at the focal plane of the lens in order to select the $+1$ diffraction order where the states were prepared. Note that this BS is not required for the ME measurement;  we used it to assist the characterization of the prepared states \cite{Supplement}. In both arms intensity measurements were carried out with CMOS cameras (Thorlabs DCC1545M). The ME measurement was performed in the transmitted arm (dashed box), where the camera was placed at the focal plane of the lens, right behind the slit diaphragm; in the reflected arm, the light distribution was imaged onto the other camera \cite{Supplement}. The SLM and both cameras were connected to a computer that controlled the former and collected and stored data from the latter. (b) Schematic of the state preparation and ME measurement stages (see text for details).}
\end{figure}

\begin{figure*}[tbp]
\centerline{\includegraphics[width=1\textwidth]{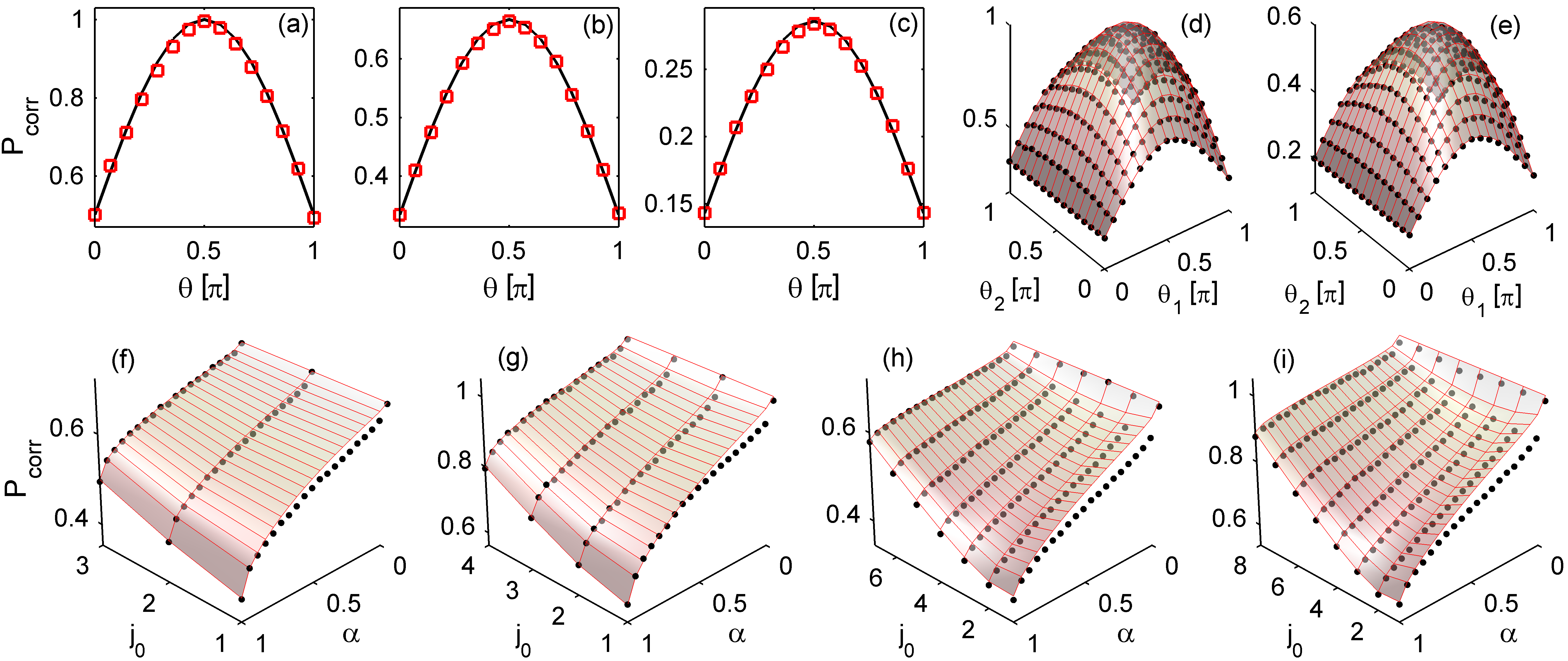}}
\caption{\label{fig:results} Optimal probability of correctly identifying the states $P_{\rm corr}$ [Eq.~(\ref{eq:Pcorr})] as a function of their coefficient parameter(s) (see text). Experimental results (squares and points) and theoretical predictions (solid lines and surfaces) of ME discrimination for $N$ states in dimension $D$ ($N$ $\times$ $D$): (a) $2\times 2$, (b) $3\times 2$, (c) $7\times 2$, (d) $3\times 3$, (e) $5\times 3$, (f) $6\times 4$, (g) $5\times 5$, (h) $12\times 8$,  (i) $9\times 9$. }
\end{figure*}

The experimental setup used to optically demonstrate the ME discrimination among symmetric states is shown and described in Fig.~\ref{fig:setup}(a). We use a bright beam from a single-mode laser as our light source, which is usual in most optical implementations of quantum-state discrimination \cite{Barnett97,Clarke01,Mohseni04,Mosley06}.  The transverse spatial profile of this beam is proportional to the transverse probability amplitude of a single-photon multimode field. With the procedure described next, it will mimic the quantum state (\ref{eq:sym-states}) of a single-photon qudit. The arrangement within the light shaded region enables the preparation of arbitrary pure states of spatial qudits, as recently demonstrated \cite{Solis-Prosser13}. This is possible by modulating the transverse spatial profile of the incoming field as follows: an array of $D$ rectangular slits with a blazed diffraction grating inside each one is displayed on the screen of the phase-only spatial light modulator. [Figure~\ref{fig:setup}(b) shows a typical phase mask for $D=3$.] In the back focal plane of a converging lens, the beam portions impinging within the slit zones are diffracted into different orders ($0,\pm 1,\ldots$); the portions impinging outside the slit zones go to the zeroth diffraction order. If we choose one of the high orders, say $+1$, to prepare the states, the amplitude of their complex coefficients is obtained by controlling the amount of light diffracted by each slit to that order, which is a function of the phase modulation depth of the gratings. The relative phases are set either by adding a constant phase value to the gratings \cite{Solis-Prosser13} or by lateral displacements of them \cite{Varga14} (here we interchanged between these two procedures \cite{Supplement}). Finally, a slit diaphragm filters out the $+1$ diffraction order from the others and the emerging light will be a coherent superposition of the which-slit modes $\{|n\rangle\}_{n=0}^{D-1}$ modulated by proper complex coefficients, thus representing the desired qudit state of Eq.~(\ref{eq:sym-states}). A schematic of the preparation stage is shown in Fig.~\ref{fig:setup}(b). The inset presents an intensity  measurement at the image plane for a state $|\psi_j\rangle$ prepared with the phase mask shown there.

The ME measurement is performed through spatial postselection. Let $\hat{\Pi}(x,z)=|\mu(x,z)\rangle\langle\mu(x,z)|$ denote the measurement operator associated with a pointlike detector in the transverse position $x$ and a longitudinal distance $z\in[f,2f]$ from a  converging lens of focal length $f$. In an $N$-dimensional space this detector postselects the state $|\mu(x,z)\rangle\propto\sum_{\ell=0}^{N-1}\varphi_\ell(x,z)|\ell\rangle$, where the complex coefficients $\varphi_\ell(x,z)$ are given by the Fresnel diffraction integral of  slit $\ell$ calculated in $(x,z)$ \cite{Solis-Prosser11}. If the detector is placed at the focal plane of the lens ($z=f$), it will postselect the state $|\mu(x,f)\rangle=\frac{1}{\sqrt{N}}\sum_{\ell=0}^{N-1}\omega^{xdN\ell /\lambda f}|\ell\rangle$, where $d$ is the slit separation and $\lambda$ is the light wavelength. Now, consider an array of $N$ detectors at $z=f$ distributed along the transverse positions $x_k=-\lambda fm_k/dN$, where $k=0,\ldots,N-1$, and $m_k=k$ if $0\leq k\leq N/2$ or $m_k=k-N$ if $N/2< k\leq N-1$ \cite{Supplement}. It is easy to see that each detector in this array postselects the state $|\mu(x_k,f)\rangle=\hat{\mathcal{F}}_N|k\rangle$. From Eq.~(\ref{eq:Fourier}), we have
\begin{equation}       \label{eq:Npos}
\hat{\Pi}(x_k,f)\equiv\hat{\Pi}^{\rm ME}_k,\;\;\;\;{\rm for}\;\;\;\;x_k=-\lambda fm_k/dN.
\end{equation}
Therefore, with such a detection scheme we can implement the ME discrimination of symmetric states. From the discussion following Eq.~(\ref{eq:Fourier}), for $N>D$, the $D$-slit array can be viewed as an array of $N$ slits where $D$ input modes are used to encode the states and the remaining $N-D$ modes are in their respective vacuum states. The propagation through the lens system provides the unitary coupling between these two subspaces and the projective measurement (\ref{eq:Npos}) accomplishes the optimal POVM.  

\begin{figure*}[tbp]
\centerline{\includegraphics[width=1\textwidth]{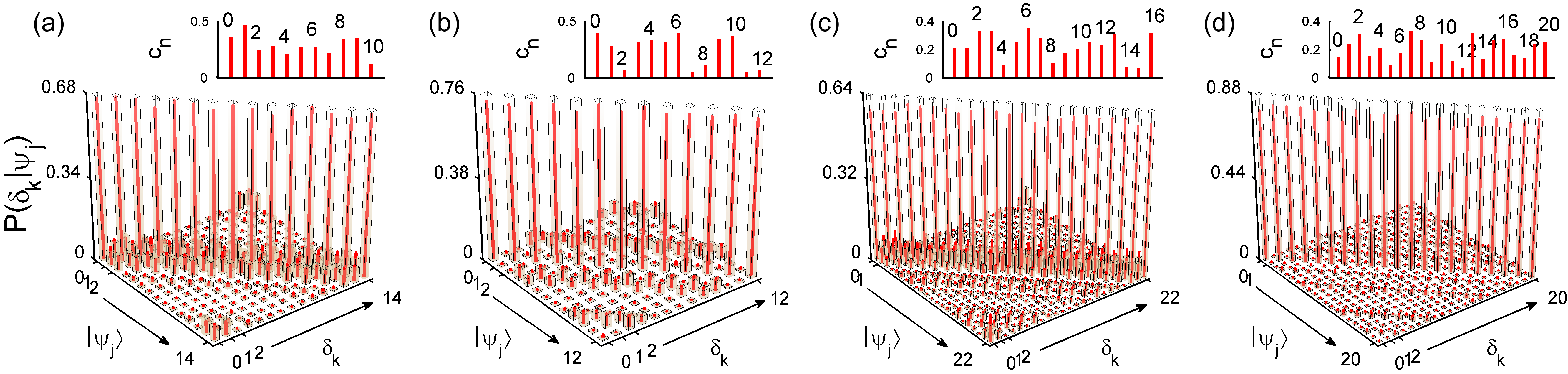}}
\caption{\label{fig:results2} Probabilities $P(\delta_k|\psi_j)$ in the ME measurement (\ref{eq:Fourier}) for a set of states defined by the coefficient amplitudes $\{c_n\}_{n=0}^{D-1}$ shown in the inset. Experimental results (red stems) and theoretical predictions (empty bars) for discrimination among $N$ states in dimension $D$ ($N$ $\times$ $D$): (a) $15\times 11$, (b) $13\times 13$, (c) $23\times 17$, (d) $21\times 21$.} 
\end{figure*}

In our setup, the ME measurement is shown in the dashed box of Fig.~\ref{fig:setup}(a). Figure~\ref{fig:setup}(b) shows a schematic of this stage with an example of $D=3$ and $N=5$. Note that we used the same lens to assist the preparation and measurement stages described above. This was done to simplify the setup, as our main goal was to demonstrate the optimal discrimination process. The same results would be achieved if the detection were performed after letting the spatial qudit state propagate through an arrangement of lenses taking the Fourier transform of the which-slit modes (possibly having to rescale the $x_k$'s).

Our implementation comprised both linearly independent and dependent states in every dimension from $D=2$ to $D=21$. Their coefficient amplitudes,  $\{c_n\}_{n=0}^{D-1}$ in Eq.~(\ref{eq:sym-states}), were parametrized as follows. For $D=2$, $c_0=\cos\frac{\theta}{2}$ with $\theta\in[0,\pi]$; for $D=3$, $c_0=\sin\frac{\theta_1}{2}\cos\frac{\theta_2}{2}$ and $c_1=\sin\frac{\theta_1}{2}\sin\frac{\theta_2}{2}$ with  $\theta_{1(2)}\in[0,\pi]$. These hyperspherical coordinates enabled the experiment to be performed for arbitrary symmetric states in those dimensions. However, for $D\geq 4$, this approach would prevent us to represent the results in a single plot. Thus, for $D=4$ to 9 we defined $c_n^2\propto 1-\sqrt[D]{\frac{n-j_0+1}{D-j_0}\alpha}$ if $n\geq j_0$, 
and $c_n^2\propto 1$ if $n<j_0$, where $j_0=1,\dots,D-1$ and $\alpha\in[0,1]$. Using only two parameters $(j_0,\alpha)$ we were able to test a large diversity of sets of symmetric states in those dimensions and, at the same time, to obtain surfaces for the probabilities in Eq.~(\ref{eq:Pcorr}) with good contrast \cite{Supplement}. This latter aspect ensured that the tested sets were well distinguishable from one another, since $P_{\rm corr}$ in the ME discrimination is also a measure of distinguishability among sets of states \cite{Chefles02}. As $D$ gets larger, to experimentally build up those surfaces became very time consuming. Thus, for $D=10$ to 21 we found it sufficient to generate a few sets of states  (three for each $D$) to perform the discrimination. In order to avoid any bias in our choice, the coefficients were randomly selected \cite{Comment}. 

With the preparation and measurement stages outlined above, the experiment is carried out in the following way: we first define the set of symmetric states by specifying $D$, $N$, and the coefficient amplitudes $\{c_n\}_{n=0}^{D-1}$ [Eq.~(\ref{eq:sym-states})]. Afterwards, we prepare one state of this set and perform the discrimination process on it by measuring, with a CMOS camera, the light intensity at the $N$ transverse positions defined in Eq.~(\ref{eq:Npos}). This is exemplified in Fig.~\ref{fig:setup}(b). The measured intensities are integrated over $y$ [this direction, also shown in Fig.~\ref{fig:setup}(b), is not relevant for us], the background noise is subtracted, and a small compensation for the detection efficiency due to diffraction is applied \cite{Supplement}. These procedures are repeated for each state in the set. Denoting $I_{kj}$ as the resulting intensity at detector $k$ when the input state is $|\psi_j\rangle$, the probability of identifying it (correctly or incorrectly) is obtained as $[P(\delta_k|\psi_j)]_{\rm expt}=I_{kj}/\sum_{\ell=0}^{N-1}I_{\ell j}$, from which we estimate the overall probability of correct identification, $P_{\rm corr}$, using the expression in the middle of Eq.~(\ref{eq:Pcorr}). 

In total, the experiment was performed for 1851 sets of symmetric states, comprising nearly 14000 different states \cite{Supplement}. Figures~\ref{fig:results} and \ref{fig:results2} show a collection of our results for the values of $N$ and $D$ specified there. In Fig.~\ref{fig:results} we plot $P_{\rm corr}$ as a function of the parameter(s) of the state coefficients defined earlier. The experimental results are given by the squares and points while the theoretical predictions---computed from the rightmost side of Eq.~(\ref{eq:Pcorr})---by the solid curves and surfaces. The error bars in these graphs were smaller than the size of the data points. In Fig.~\ref{fig:results2} we plot $P(\delta_k|\psi_j)$, defined above Eq.~(\ref{eq:Pcorr}). Each graph shows these probabilities for a single set of states settled by the coefficient amplitudes shown in the inset. The red stems are the experimental results and the empty bars the expected theoretical values given by $[P(\delta_k|\psi_j)]_{\rm theo}=|\langle\mu_k|\psi_j\rangle|^2$. We observe in both figures the close agreement between theory and experiment. 

\begin{figure}[b]
\centerline{\includegraphics[width=.48\textwidth]{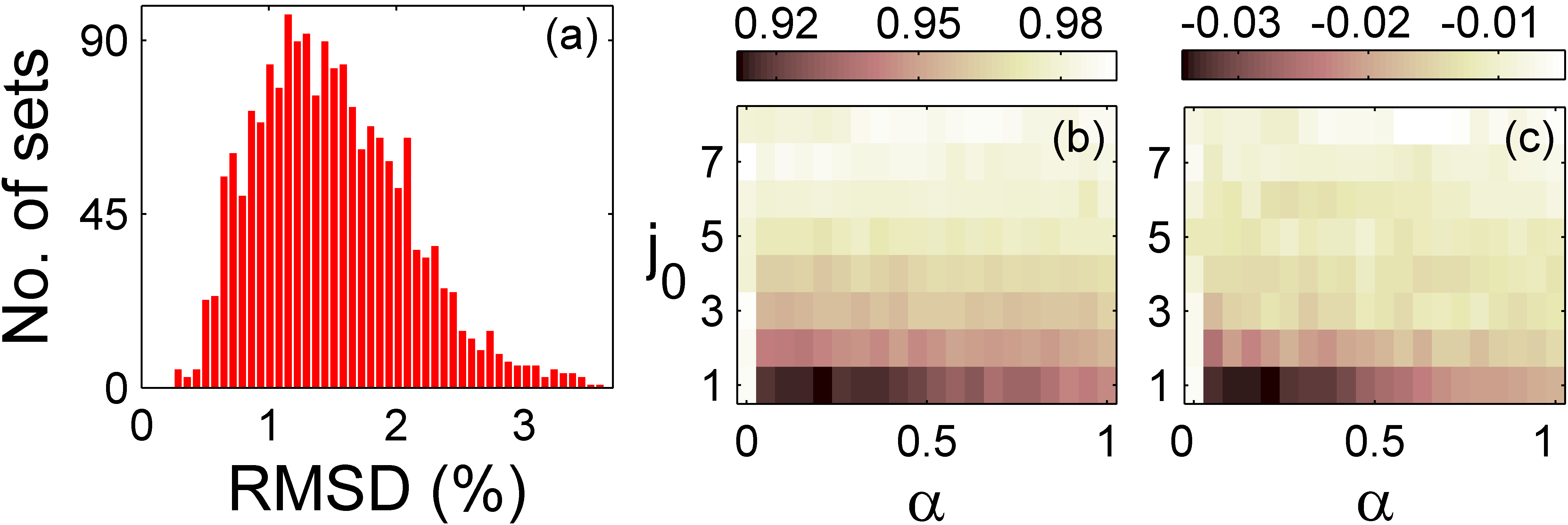}}
\caption{\label{fig:errors} (a) Histogram with the distribution of the root-mean-square deviations of the experimental results from the theoretical values. (b) Average fidelities of state preparation and (c) RMSD$\times(-1)$ for the results shown in Fig.~\ref{fig:results}(i). }
\end{figure}

For each of the tested sets we calculated the root-mean-square deviation (RMSD) of the results from the optimal theoretical values. With the residuals defined by $R_{kj}\equiv[P(\delta_k|\psi_j)]_{\rm theo}-[P(\delta_k|\psi_j)]_{\rm expt}$, we have ${\rm RMSD}=\sqrt{\sum_{j,k=0}^{N-1}R_{kj}^2/N^2}$. Figure~\ref{fig:errors}(a) shows a histogram of the obtained deviations, which ranged from $0.3\%$ to 3.6\%. For 80\% of the sets the RMSDs were below 2\%. The errors in our experiment are caused mainly by aberrations of the optical elements and the imperfect preparation of the input states, which is, possibly,  responsible for the largest deviations. For instance, examining the results of Fig.~\ref{fig:results}(i), it can be seen that the worse ones occurred for $j_0=1,2$ and some values of $\alpha\neq 0$. For this particular realization, we plot in Figs.~\ref{fig:errors}(b) and \ref{fig:errors}(c) the average fidelities of preparation for the states in each set and the RMSD$\times(-1)$, respectively. It is clearly observed that the largest deviations from the theoretical predictions correspond to the sets of states with lower fidelities. Despite all this, the deviations in our experiment, as shown in Fig.~\ref{fig:errors}(a), were considerably small taking into account that experimental implementations are always subjected to errors, the wide variety of states tested, and the high dimensions we have explored. Therefore, it is safe to say that our scheme is indeed optimal. 

In conclusion, we have presented the first experimental demonstration of minimum-error discrimination among nonorthogonal states of a $D$-dimensional qudit. Our measurement scheme was designed to distinguish symmetric pure states,  an important resource in quantum information \cite{Chefles98,Croke06,Jimenez11,Sugimoto12,Herzog12,Nakahira12,Ban97,Phoenix00,Neves12,Prosser14,Pati05}. Envisaging its application to quantum protocols, a single photon implementation of this experiment would require a source with a sufficiently large transverse spatial coherence, in order to prepare high-quality states. The preparation and measurement stages would be exactly the same described earlier. The detector array, however, will be required to have single-photon counting capability. Some candidates for this include charged-coupled device cameras (CCDs), either an intensified CCD \cite{Fickler13} or an electron multiplying CCD \cite{Zhang09}, and also single-photon avalanche diode arrays based on CMOS technology \cite{Scarcella13,Unter16}.

The ME measurement is central not only to many applications in quantum information, but also to put forward the implementation of ``complete'' probabilistic discrimination protocols in high dimensions: by iterating the discrimination process in case of failed attempts \cite{Neves12}, one can significantly increase the information gain about the input states. 
Our demonstration constitutes a building block for future realizations of these protocols and, accordingly, will benefit a variety of tasks in high-dimensional quantum information processing \cite{Phoenix00,Neves12,Prosser14,Pati05}.

\begin{acknowledgments}
This work was supported by the Brazilian agencies CNPq (Grant No.~485401/2013-4), FAPEMIG (Grant No.~APQ-00149-13), and PRPq/UFMG (Grant No.~ADRC-01/2013). M.A.S-P., O.J., and A.D. acknowledge financial support from  Millennium Scientific Initiative (Grant No.~RC130001) and FONDECyT (Grants No.~1140635 and No.~11121318).
\end{acknowledgments}

\input{Suppl_mat.tex}

\clearpage

\end{document}

%% file: Suppl_mat.tex
%
%

\appendix
%
%
%
%
%
%
%

\maketitle

\section{Preparation and characterization of symmetric states}

\subsection{State preparation and phase encoding}

\par Let us consider a pure qudit state $|\psi\rangle=\sum_{\ell=0}^{D-1}c_\ell|\ell\rangle$, where $\sum_{\ell=0}^{D-1}|c_\ell|^2=1$. This state can be encoded onto the one-dimensional transverse spatial modes of an optical field by modulating its transverse profile with a transmission function $t(x,y)$ given by
\begin{align}
	t(x,y) = {\rm rect}\left(\frac{y}{L}\right) \sum_{\ell=0}^{D-1} t_{\ell}~{\rm rect}\left( \frac{x-\xi_{\ell} d}{2a} \right),
\end{align}
where $\xi_{\ell}=\ell-(D-1)/2$, $d$ is the distance between the centers of two neighboring slits, $2a$ is the width of each slit, $L$ is the slit length, ${\rm rect}(\bullet)$ are rectangle functions, and $t_\ell$ is the transmission coefficient of each slit, defined as 
\begin{align*}
t_\ell=&\frac{c_\ell}{c_{\max}},\\
c_{\max}=&\max\{|c_\ell|,\ell=0,\dots,D-1\}.
\end{align*}

\par The procedures indicated in Refs.~\cite{Solis-Prosser13,Varga14}, based on the use of diffraction gratings generated by a phase-only spatial light modulator (SLM) and the selection of its first diffraction order, were followed in order to generate the states to be discriminated. Briefly, the modulus of the transmission function $|t(x,y)|$ is encoded by a diffraction grating with variable phase modulation depth. Afterwards, the complex phase of the transmission function, ${\rm Arg}\bm{(}t(x,y)\bm{)}$, is encoded either by adding the phase directly to the grating \cite{Solis-Prosser13} or by displacing the grating by a certain amount of pixels for each slit \cite{Varga14}. These methods are known as phase addition (PA) and grating displacement (GD), respectively. The aforementioned procedures were complemented with a convenient choice of the grating period and orientation~\cite{Marquez07} and an appropriate configuration of the SLM---i.e. phase modulation response to the addressed gray levels~\cite{Moreno04} and digital electric addressing sequences~\cite{Lizana10}---in order to improve the efficiency and temporal stability of the modulated phase.

\begin{figure}[!thbp]
\centerline{\includegraphics[height=0.58\textheight]{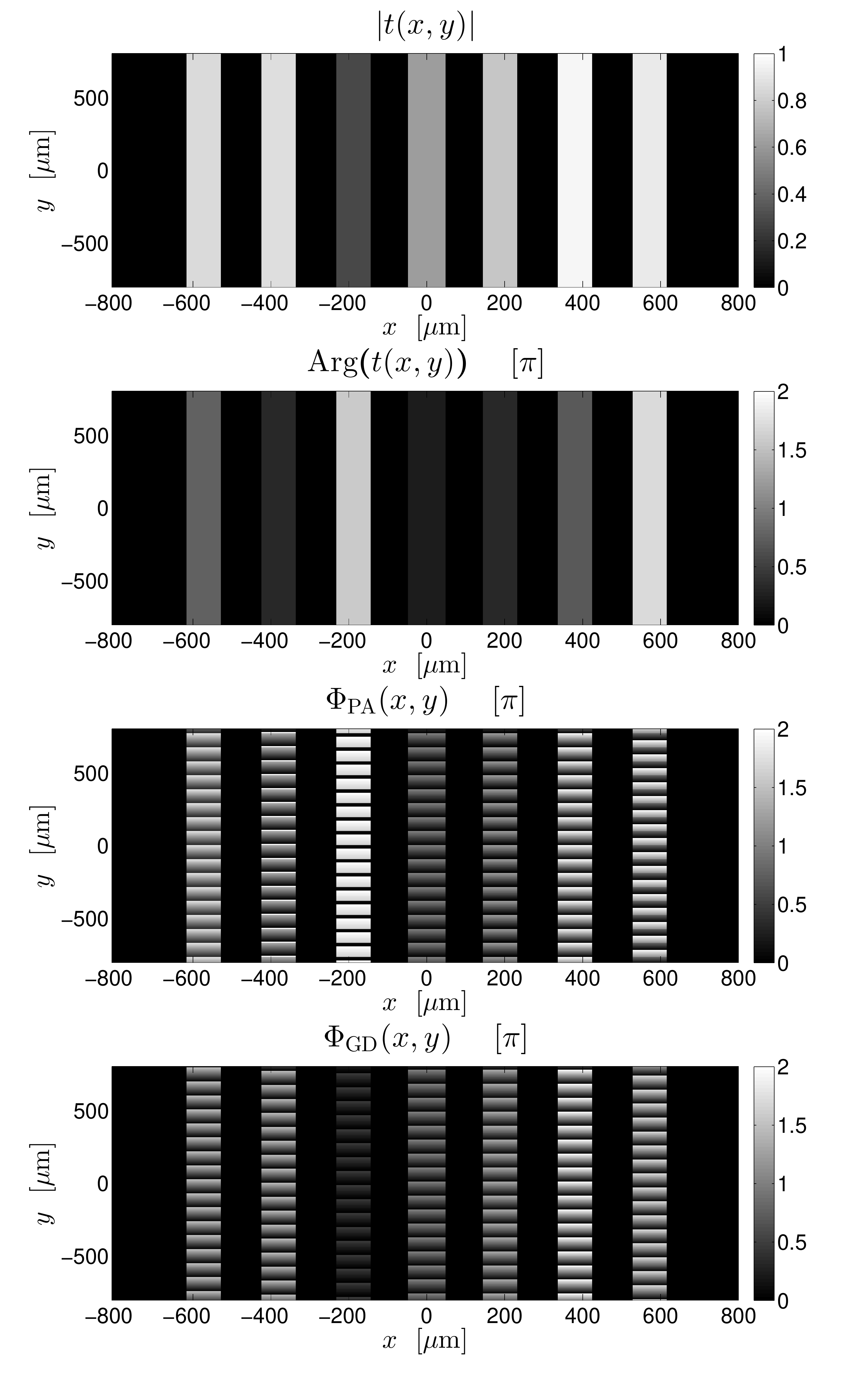}}
\caption{\label{fig:Masks} An illustrative example of the state generation method for spatial qudits used in this work. For a given state $|\psi\rangle$, a transmission function $t(x,y)$ must be generated, whose modulus $|t(x,y)|$ and ${\rm Arg}\bm{(}t(x,y)\bm{)}$ are shown in the first two pictures from top to bottom, respectively. Depending on the method being used, the phase mask able to implement the state is given by $\Phi_{\rm PA}(x,y)$   if the phase addition method is selected (third figure). Otherwise, $\Phi_{\rm GD}(x,y)$ stands for the phase mask using the grating displacement method (fourth figure). }
\end{figure}

\par Figure~\ref{fig:Masks} shows an example of $|t(x,y)|$, ${\rm Arg}\bm{(}t(x,y)\bm{)}$ and the corresponding phase masks addressed at the SLM (via PA or GD method) used to prepare a given spatial qudit state with $D=7$. States ranging from $D=2$ to $D=9$ were generated using the PA method, whereas GD method was used for $D=10$ to $D=21$. This choice was motivated by the quality of the states being generated, whose estimation is explained in Sections~\ref{sec:tomoqubits} and~\ref{sec:tomoqudits}.

\subsection{State characterization for qubits \label{sec:tomoqubits} }

\par
\begin{figure}[tbp]
\centerline{\includegraphics[width=.49\textwidth]{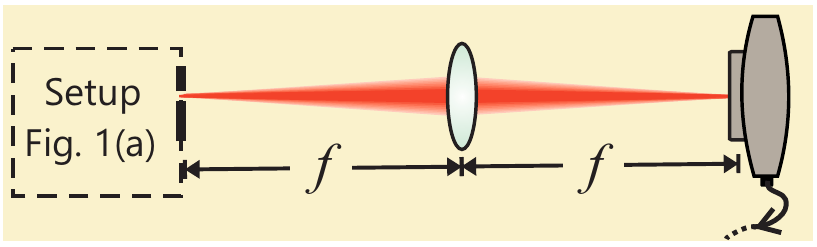}}
\caption{\label{fig:setupSM} An additional CMOS camera is located at the image plane of the SLM [Figure~1(b) of the main text] in order to characterize the state preparation procedure. Although recommendable, this stage is not required for the discrimination process.}
\end{figure}

 For a spatial qubit, a complete tomography can be performed since an informationally complete set of measurements can be taken from four transverse positions of the detectors in the Fourier plane and two ones located at the image plane \cite{Solis-Prosser11}. For this reason, the optical setup included a CMOS camera situated at the image plane of the SLM in order to perform the state estimation, as Fig.~\ref{fig:setupSM} depicts. These six measurements complete a set of three orthogonal bases. 

\par  The first basis, corresponding to the states $|b_{0j}\rangle=(|0\rangle+e^{i j\pi}|1\rangle)/\sqrt{2}$ with $j=0,1$, can be measured at positions $x_{0j}=j\pi f/kd$ in the focal plane. The second basis contains the states $|b_{1j}\rangle=(|0\rangle+ie^{i j\pi}|1\rangle)/\sqrt{2}$ and can be measured at $x_{1j}=(j-\tfrac{1}{2})\pi f/2kd$, also in the focal plane. The third basis is the computational basis $|b_{2j}\rangle=|j\rangle$, which is measured at positions $x_{2j}=-(j-\tfrac{1}{2})d$ in the image plane (the minus sign at the beginning is consequence of the inversion of the image created by any lens system). 

\par Once the intensities were recorded, the probabilities are obtained by dividing each intensity by the total intensity within each set, i.e., $p_{\mu j} = I_{\mu j}/(I_{\mu 0}+I_{\mu 1})$. The first estimation of the  state is given by \cite{Ivanovic81}
\begin{align}
	\hat{\rho}_{\rm tomo} = \sum_{\mu=0}^{2} \sum_{j=0}^{1} p_{\mu j}|b_{\mu j}\rangle\langle b_{\mu j}| - \hat{I},\label{tomo_d2}
\end{align}
where $\hat{I}$ is the identity operator on a two-dimensional Hilbert space. This method for state reconstruction relies on measurements performed on three mutually unbiased bases (MUBs) and, hence, it is commonly referred to as MUB tomography.

\par The operator $\hat{\rho}_{\rm tomo}$ defined in Eq.~(\ref{tomo_d2}) is, by construction, an hermitian matrix. However, due to experimental imperfections, this matrix is not always positive definite and, hence, it might not represent a physical state. The same measurements taken can be used for a second procedure known as maximum likelihood estimation (MLE). This procedure is an optimization problem in which $\hat{\rho}_{\rm tomo}$  can be used as a starting point for the numerical optimization in order to obtain the appropriate density operator which most likely describes the experimental data \cite{James01}. This second estimated state, defined as $\hat{\rho}_{\rm MLE}$, is used afterwards to compute its fidelity with respect to the target state $|\psi\rangle$, given by $F(|\psi\rangle,\hat{\rho}_{\rm MLE})= \langle\psi|\hat{\rho}_{\rm MLE}|\psi\rangle$, in order to characterize the quality of the state preparation.

\subsection{State characterization for qudits \label{sec:tomoqudits} }

\begin{figure*}[!tbp]
\centerline{\includegraphics[width=.95\textwidth]{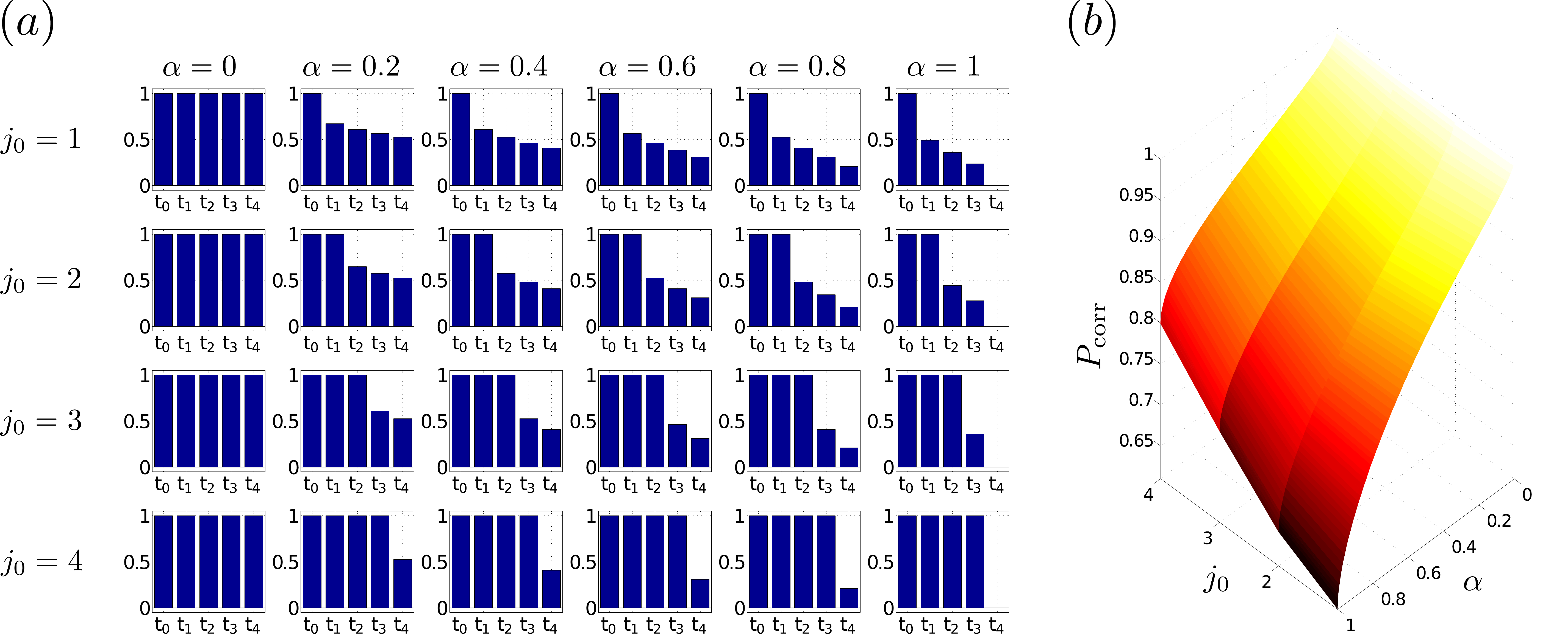}}
\caption{\label{fig:Cascada} $(a)$ Simplified depiction of the transmission coefficients $t_k$ of each slit used to prepare the coefficients $c_k$ as a function of $j_0$ and $\alpha$ for $D=5$. $(b)$ Optimal minimum error probability as a function of $\alpha$ and $j_0$ for $N=D=5$.}
\end{figure*}

\par Unlike the case of spatial qubits, an arbitrary $D$-dimensional (${D>2}$) qudit state cannot be properly reconstructed only from measurements at the Fourier and image planes of a single imaging system and such a task requires the assistance of an additional SLM \cite{Lima11}. Nevertheless, the states generated in our setup are nearly pure due to the high transverse coherence length of the laser beam, which is attested by the high visibility of the interference patterns. Thus, an estimation can be performed by using least squares fitting as follows. We first write the state to be estimated as $|\psi_{\rm est}\rangle=\sum_{j=0}^{D-1}a_j|j\rangle$, where $|j\rangle$ is the state of the transverse optical mode associated with $j$th slit. The intensities $I_j$ of each slit at the image plane allows us to obtain the complex modulus of each coefficient $a_j$, according to $|a_j|^2=I_j/(\sum_{k=0}^{D-1}I_k)$. This allows to rewrite the estimated state as $|\psi_{\rm est}\rangle\propto\sum_{j=0}^{D-1}\sqrt{I_j}e^{i\phi_j}|j\rangle$. The expected far field intensity distribution for this state is given by
\begin{align}
	[\mathcal{I}_{\rm est}&\bm{(}\{\phi_k\},\mathcal{I}_{\max}\bm{)}]_\kappa \nonumber \\ &= \mathcal{I}_{\max}\left| \sum_{j=0}^{D-1}\sqrt{I_j}e^{i\phi_j}e^{ikjdx_\kappa/f}{\rm sinc}\left(\frac{kax_{\kappa}}{f}\right)   \right|^2,
\end{align}
for every recorded transverse position $x_\kappa$ in the focal plane, where $\mathcal{I}_{\max}$ is a global proportionality constant for each experimental intensity distribution, and ${\rm sinc}(z)=\sin(z)/z$. Then, the phase retrieval is performed from the experimentally obtained far field intensity distribution $[\mathcal{I}_{\rm exp}]_\kappa$ by least squares according to

\begin{align}
\Delta\bm{(}\{\phi_k^{\star}\},\mathcal{I}_{\max}^{\star}\bm{)} =& \min_{\{\phi_k\},\mathcal{I}_{\max}} \Delta\bm{(}\{\phi_k\},\mathcal{I}_{\max}\bm{)}, \label{lsq}
\end{align}
where 
\begin{align}
\Delta\bm{(}\{\phi_k\},\mathcal{I}_{\max}\bm{)} = & \sum_{\kappa} \bm{\bigg(}[\mathcal{I}_{\rm est}\bm{(}\{\phi_k\},\mathcal{I}_{\max}\bm{)}]_\kappa -   [\mathcal{I}_{\rm exp}]_\kappa \bm{\bigg)}^2,
\end{align}
and $\{\phi_k^{\star}\}$ and $\mathcal{I}_{\max}^{\star}$ are the parameters that allow to minimize $\Delta\bm{(}\{\phi_k\},\mathcal{I}_{\max}\bm{)}$. Once the minimization has been carried out, the estimated state is completed by the phases that solve the problem of Eq.~(\ref{lsq}), i.e.,
\begin{align}
	|\psi_{\rm est}\rangle\propto\sum_{j=0}^{D-1}\sqrt{I_j}e^{i\phi_j^{\star}}|j\rangle.
\end{align}
This state is then used to characterize the quality of the preparation through the fidelity $F(|\psi_{\rm est}\rangle,|\psi\rangle)=|\langle\psi_{\rm est}|\psi\rangle|^2$  between the estimated state and the target state $|\psi\rangle$ intended to be prepared.

\subsection{State parametrization for $D=4$ to $D=9$}

\par In general, we studied the case of discrimination among $N$ states of the form 
\begin{align}
	|\psi_j\rangle = \sum_{k=0}^{D-1} c_k \omega^{jk}|k\rangle,~~~~~~j\in\left\{0,\cdots,N-1\right\},\label{eq:psij}
\end{align}
where $c_k$ are normalized non-negative coefficients and $\omega=\exp({2\pi i/N})$. Coefficients $c_k$ can be easily parametrized by using hyperspherical coordinates. In dimension~2, the Bloch parametrization is recovered, as $|\psi_j\rangle= {\cos(\theta/2)|0\rangle} {+\omega^{j}\sin(\theta/2)|1\rangle}$, where $\theta\in [0,\pi]$. The three dimensional case is easily parametrized as well: $|\psi_j\rangle= \sin(\theta_1/2)\cos(\theta_2/2)|0\rangle +\omega^{j}\sin(\theta_1/2)\sin(\theta_2/2)|1\rangle +\omega^{2j}\cos(\theta_1/2)|2\rangle$, with $\theta_{1,2}\in [0,\pi]$. And so on for larger dimensions. For each dimension $D$, we require $D-1$ angles $\theta_m\in [0,\pi]$ in order to be able of preparing arbitrary states of the form given by Eq.~(\ref{eq:psij}). 
\par However, for $D\geq 4$, a practical problem arises: It is not possible to represent the results in a single plot as more than 2 angles are required. Additionally, the involved times increase exponentially since, considering $M$ as the number of values taken of each angle $\theta_m$, a complete discrimination of $N$ states for \emph{all} fiducial states needs a number of measurements equal to $N\times M^{D-1}$, becoming extremely time-consuming for large dimensions. 
\par In order to overcome this drawback, we devised a strategy in which these $D$-dimensional fiducial states are parametrized by only two parameters, making the graphic representations of the results feasible. The challenge lies in the way the parametrization is chosen, as we imposed that (i) it must provide a reasonable diversity of fiducial states, and (ii) the surface for the plot of the theoretical minimum-error probability must have a good contrast, i.e., a reasonable difference between the minimum and maximum expected values. This last requirement, beyond aesthetic reasons, is motivated by the fact that the ME discrimination probability can also be used as a distinguishability measure within a set of quantum states \cite{CheflesDist}: A surface exhibiting few changes and no slopes could have been interpreted as having measured sets with very similar distinguishabilities. Hence, the experiments would have considered a very particular subset of states from the Hilbert space rather than being a proof-of-principle of the general applicability of this method for sets constructed from arbitrary fiducial states. We defined the following biparametric representation of the transmission coefficients 
\begin{align}
t_j= \begin{cases}
	\displaystyle\sqrt{1-\sqrt[D]{\frac{j-j_0+1}{D-j_0}\alpha}} ~~~~&~{\rm if}~~j\geq j_0\\
	{\,\,} & {\,\,}\\
	1 ~~~~&~{\rm if}~~j< j_0\\
\end{cases},
\end{align}
with $\alpha\in[0,1]$ and $j_0=1,\dots,D-1$, where the state coefficients computed as 
\begin{align}
	c_j = \frac{t_j}{\mathcal{N}_{\alpha,j_0}},~~~~~~\mathcal{N}_{\alpha,j_0}= \sqrt{\sum_{k=0}^{D-1}t_k^2},
\end{align}
accomplish the aforementioned requirements. The behavior of the transmission coefficients $t_j$ for each slit, where $t_j\propto |c_j|$, is depicted in Fig.~\ref{fig:Cascada} together with the expected optimal minimum error probability. As it is shown,  the transmission of the first $j_0$ slits ($0,\dots,j_0-1$) will remain unaltered as $\alpha$ increases. On the contrary, the remaining slits $(j_{0},\dots,D-1)$ will have their transmission coefficients changed as a function of $\alpha$. The measurements started, for each value of $j_0$, at $\alpha=0$, which corresponds to $N$ maximally distinguishable states in a $D$-dimensional space. It ended at $\alpha=1$, where linearly dependent states are being discriminated. For  $0<\alpha <1$, the linear dependence will only rely on the relationship between $N$ and $D$. As $\alpha$ increases, the optimal minimum error probability decreases and, therefore, distinguishability decreases as well.

\section{Detectors and data analysis\label{sec:detectors}}

\par Detectors were CMOS cameras Thorlabs DCC1545M, whose pixels are $5.2\,{\rm\mu m}$ wide squares. The camera gamma function was programmed to obtain, in terms of gray levels, a linear response with respect to the incoming optical power. The laser power and camera gain were adjusted to avoid saturation in any pixel. For every generated state, three images from the Fourier plane were captured and summed in order to obtain less noisy results. The exposure time of each image was set the closest possible to 1/60 s, which corresponds to the flicker period of the SLM. Once the data has been taken, an integration over the $y$ axis is performed in order to trace over this non relevant direction since the intensities of interest were distributed along the $x$ axis. In this way, only one-dimensional intensity distributions are recorded. The background noise was subtracted from each pattern taking the minimum of the diffraction envelope as a reference.
\par Once the one-dimensional patterns were processed, state discrimination is performed. This is done by registering the intensity at $N$ transverse positions given by $x_k=-\lambda f m_k /dN$, where $f$ is the focal length of the lens, $\lambda$ is the optical field wavelength, $N$ is the number of states being discriminated in each set, and 
\begin{align}
	m_{k} = \begin{cases} 
		k  &~{\rm if}~~~0\leq k \leq \tfrac{N}{2}\\
		k-N  &~{\rm if}~~~\tfrac{N}{2}< k \leq N-1
	\end{cases},
\end{align}
is a number that controls the distribution of the positions $x_k$ around $x=0$. We shall define the intensities $I_{k}^{(j)}$ as the intensity measured at $x_k$ when the input state is $|\psi_j\rangle$. Thus, the experimentally estimated probability of having a photodetection at the $k$th detector is given by 
\begin{align}
	P(\delta_k | \psi_j) = \dfrac{I_{kj}}{\sum_{l=0}^{N-1} I_{lj}},
\end{align}
where $I_{kj}=I_{k}^{(j)}/\eta_k$ is the intensity registered at the $k$th transverse position after being compensated by the optical diffraction efficiency, given by $\eta_k = {\rm sinc}^{2}(kax_{k}/f)$. We can note that the optical efficiency compensation depends solely on the detector position, independently of the state being discriminated. 

\par
\begin{figure*}[!t]
\centering
\includegraphics[width=0.49\textwidth]{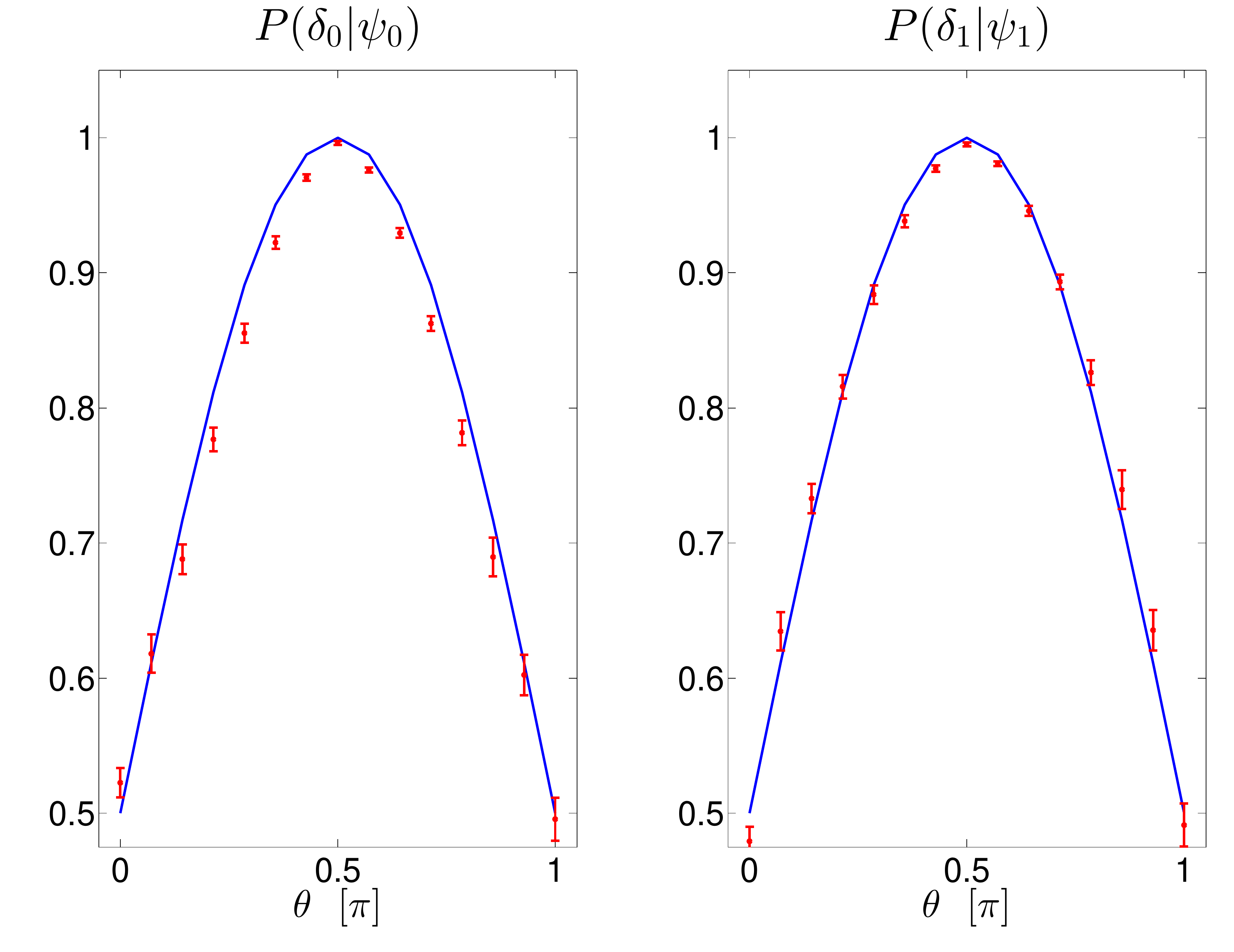}
\includegraphics[width=0.49\textwidth]{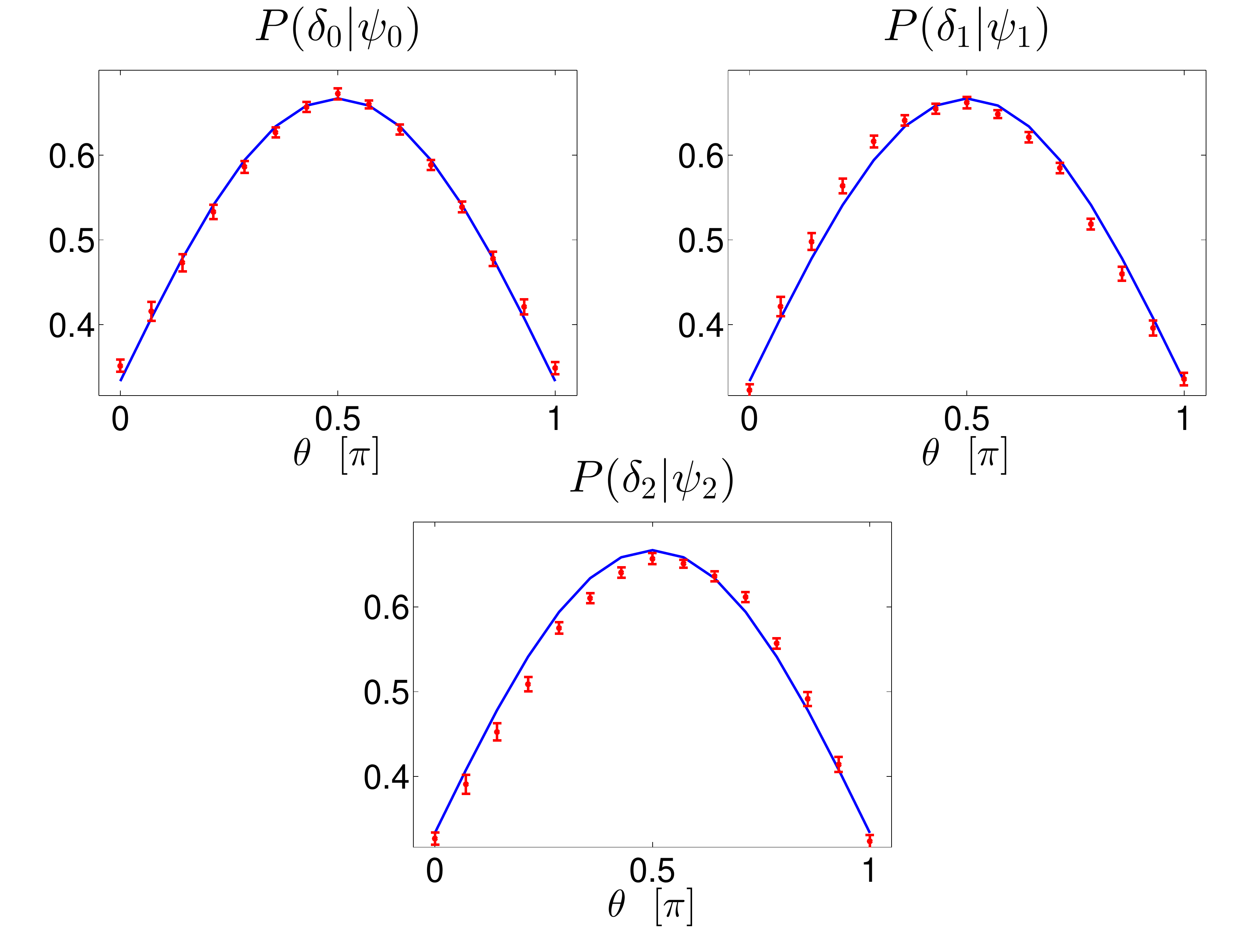}
\caption{\label{fig:D2N2} Results for state discrimination among qubit states for $N=2$ (left) and $N=3$ (right).}
\end{figure*}

\begin{figure*}[!t]
\centering
\includegraphics[height=.30\textheight]{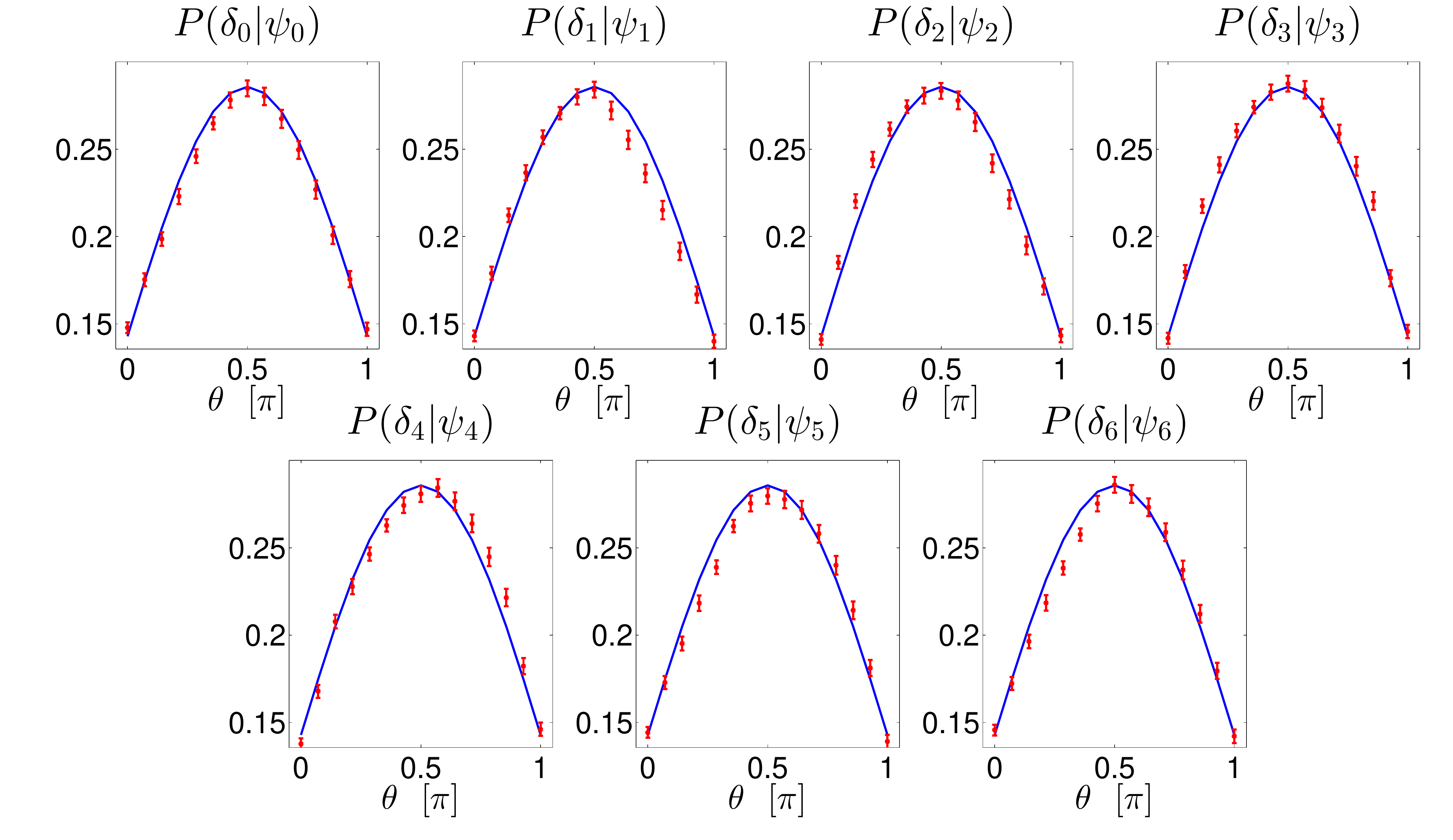}
\caption{\label{fig:D2N7} Results for state discrimination among qubit states for $N=7$.}
\end{figure*}

\begin{figure*}[!t]
\centering
\includegraphics[height=.45\textheight]{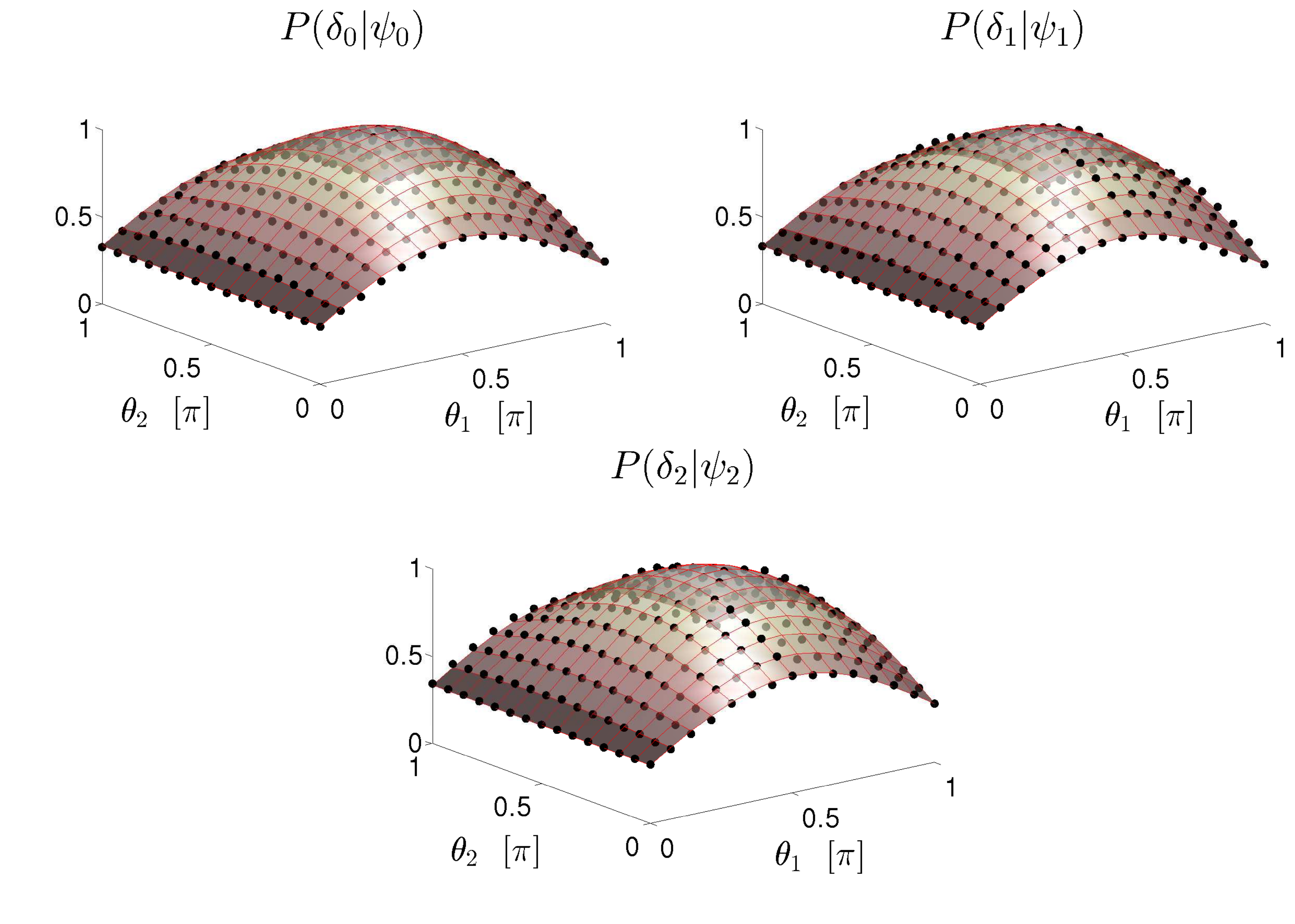}
\caption{\label{fig:D3N3} Results for state discrimination among qutrit states for $N=3$}
\end{figure*}

\begin{figure*}[!thbp]
\centering
\includegraphics[height=.45\textheight]{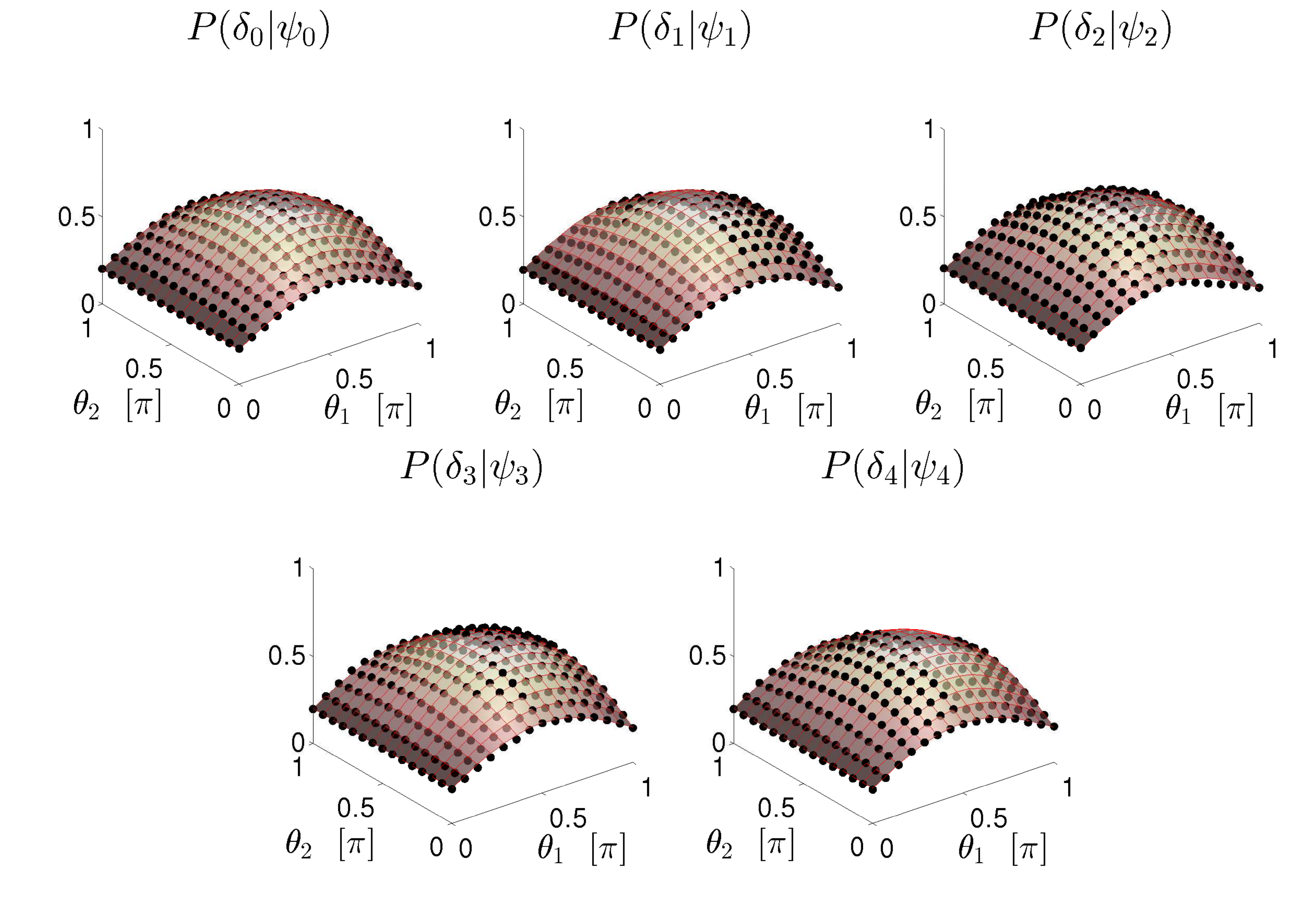}
\caption{\label{fig:D3N5} Results for state discrimination among qutrit states for $N=5$.}
\end{figure*}

\begin{figure*}[!t]
\centering
\includegraphics[height=.45\textheight]{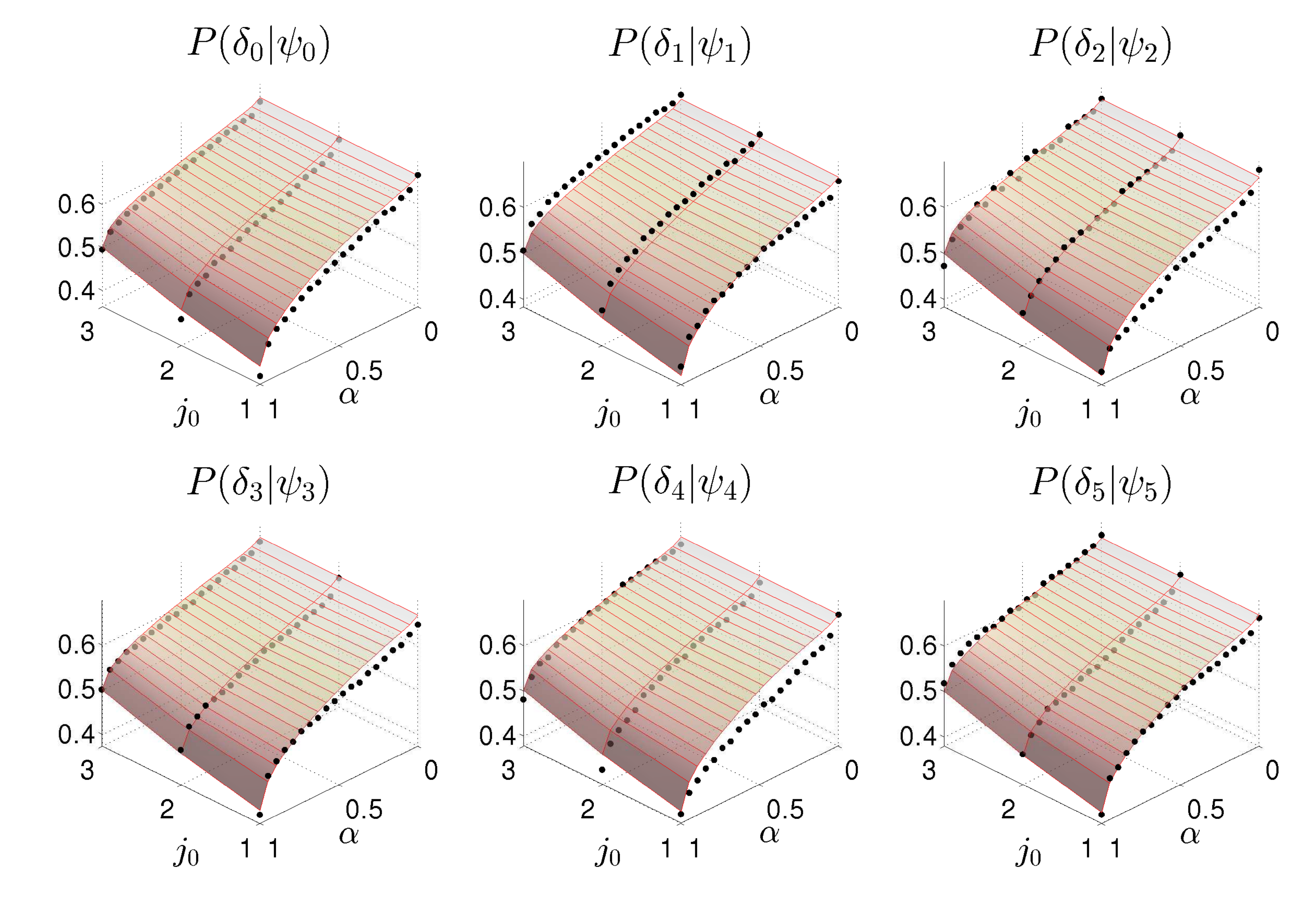}
\caption{\label{fig:D4} Results for state discrimination among qudit states for dimension 4 and $N=6$.}
\end{figure*}

 For each input state, the probability of having a correct retrodiction of the state is given by $P(\delta_j|\psi_j)$. The ME strategy aims to maximize the overall probability of correct retrodiction for each set, i.e., 
\begin{align}
	P_{\rm corr} = \dfrac{1}{N}\sum_{j=0}^{N-1}P(\delta_j|\psi_j),
\end{align}
where the factor $1/N$ corresponds to the \emph{a priori} probability of each state $|\psi_j \rangle$ being selected, which are assumed here to be equally likely. Figures~\ref{fig:D2N2} to~\ref{fig:D4} show some of the results for the probabilities of correct retrodiction $P(\delta_j|\psi_j)$ compared with their respective optimal expected value. The average $P_{\rm corr}$ for these cases has been shown in Fig.~2 of the main text.

\begin{table}[!t]
\centering
\caption{Summary of the sets of states tested.}
\label{table:sets}
\begin{tabular}{ccccc}
\hline\hline
$D$  & $N$     & Sets/$N$ & No. of sets & No. of states \\
\hline
2    & 2, 3, 7 & 15       & 45          & 180           \\
3    & 3, 5    & 225      & 450         & 1800          \\
4    & 4, 6    & 60       & 120         & 600           \\
5    & 5, 7    & 80       & 160         & 960           \\
6    & 6, 7    & 100      & 200         & 1300          \\
7    & 7, 11   & 120      & 240         & 2160          \\
8    & 8, 12   & 140      & 280         & 2800          \\
9    & 9, 13   & 160      & 320         & 3520          \\
10 to 21    & $N\geq D$   & 1      & 3         & $\sim 600$          \\
\hline
\multicolumn{3}{c}{Total} & 1851        & $\sim 13920$          \\
\hline
\end{tabular}
\end{table}

\begin{figure*}[!thbp]
\centering
\includegraphics[height=.45\textheight]{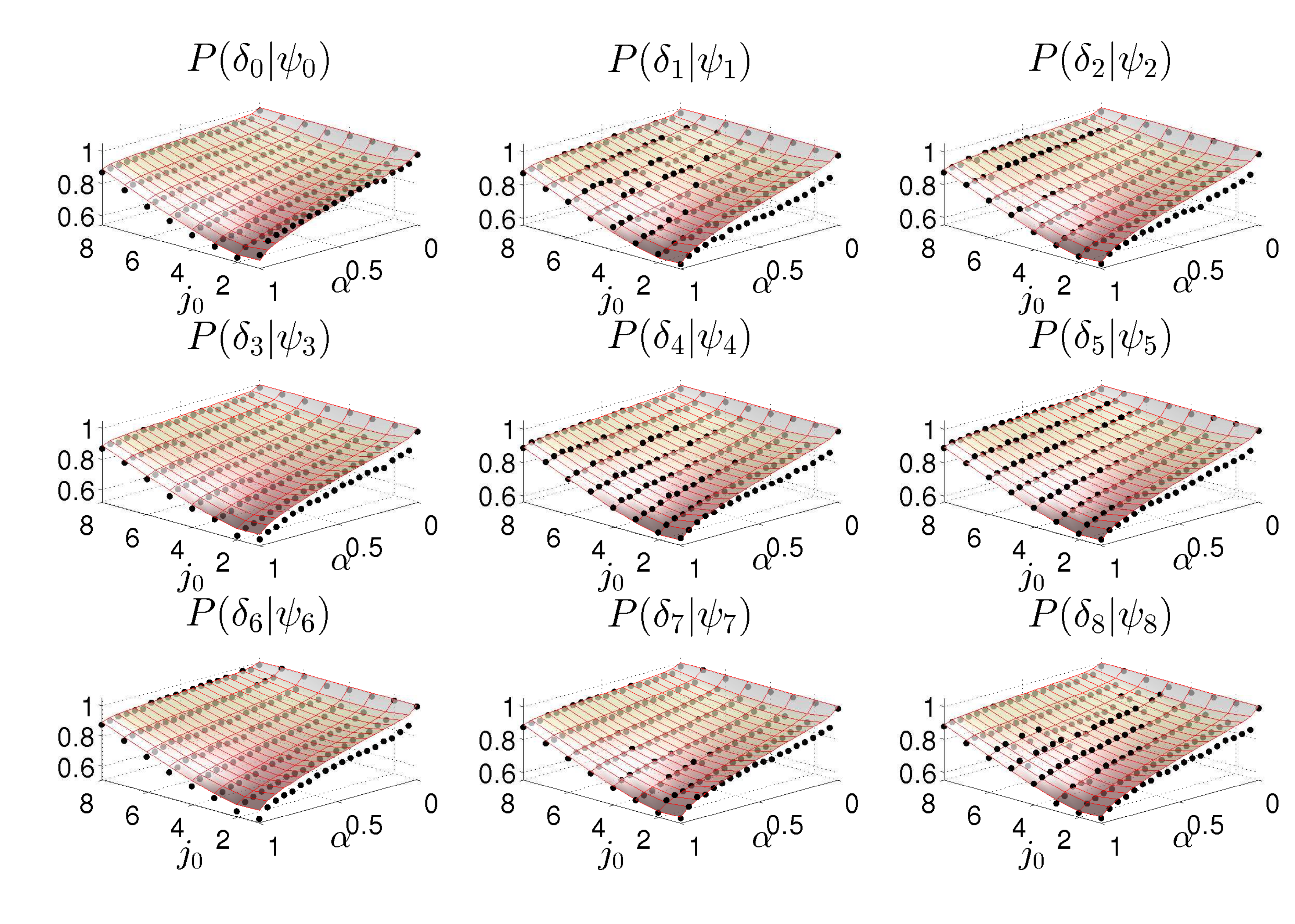}
\caption{\label{fig:D9} Results for state discrimination among qudit states for dimension 9 and $N=9$.}
\end{figure*}

\subsection{Number of sets and states tested}

\par Table~\ref{table:sets} summarizes the dimensions ($D$), number of states per set ($N$), number of sets per $N$, and the number of sets and states per $D$ tested in our experiment. A total number of 1851 sets were tested where more than 13900 states were generated and discriminated. The process of generation, characterization and discrimination of these states was completely automated through a single portable computer controlling the CMOS cameras and the SLM simultaneously. Before the experiment, a calibration of the CMOS was made in order to set the correct positions for state discrimination and state estimation. Afterwards, the full discrimination procedure was performed: For each state, a 1080$\times$1920 grayscale image containing a phase mask, like the examples $\Phi_{\rm PA}(x,y)$ and $\Phi_{\rm GD}(x,y)$ shown in Fig.~\ref{fig:Masks}, is addressed to the SLM in order to prepare the state. Then, both cameras registered the intensity distribution from Fourier and image planes. State estimation is performed according to the methods described in Secs.~\ref{sec:tomoqubits} and~\ref{sec:tomoqudits}. Discrimination procedure is described in the main text and in Sec.~\ref{sec:detectors}.

%% file: DRAFT_ME_V3.bbl
\begin{thebibliography}{99}

	\bibitem{Spekkens07} R. W. Spekkens, \href{http://dx.doi.org/10.1103/PhysRevA.75.032110}{\pra~{\bf 75}, 032110 (2007).}

	\bibitem{Leifer14} M. S. Leifer,  \href{http://dx.doi.org/10.1103/PhysRevLett.112.160404}{\prl~{\bf 112}, 160404 (2014).}

	\bibitem{Branciard14} C. Branciard, \href{http://dx.doi.org/10.1103/PhysRevLett.113.020409}{\prl~{\bf 113}, 020409 (2014).}

	\bibitem{Ringbauer15} M. Ringbauer, B. Duffus,	C. Branciard, E. G. Cavalcanti, A. G. White, and A. Fedrizzi, \href{http://dx.doi.org/10.1038/nphys3233}{Nat. Phys. {\bf 11}, 249 (2015).}

	\bibitem{Gisin02} N. Gisin, G. Ribordy, W. Tittel, and H. Zbinden, \href{http://dx.doi.org/10.1103/RevModPhys.74.145}{Rev. Mod. Phys. {\bf 74}, 145 (2002).}

	\bibitem{Grosshans03} F. Grosshans, G. Van Assche, J. Wenger, R. Brouri, N. J. Cerf, and P. Grangier, \href{http://www.nature.com/nature/journal/v421/n6920/full/nature01289.html}{Nature (London) {\bf 421}, 238 (2003)}.

	\bibitem{Reviews} For reviews of quantum-state discrimination see A. Chefles, \href{http://dx.doi.org/10.1080/00107510010002599}{Contemp. Phys. {\bf 41}, 401 (2000);} S. M. Barnett and S. Croke, \href{http://dx.doi.org/10.1364/AOP.1.000238}{Adv. Opt. Photon. {\bf 1}, 238 (2009);} J. A. Bergou, \href{http://dx.doi.org/10.1080/09500340903477756}{J. Mod. Opt. {\bf 57}, 160 (2010).}

	\bibitem{Holevo73} A. S. Holevo, \href{http://dx.doi.org/10.1016/0047-259X(73)90028-6}{J. Multivariate Anal. {\bf 3}, 337 (1973).}

	\bibitem{Yuen75} H. P. Yuen, R. S. Kennedy, and M. Lax, \href{http://dx.doi.org/10.1109/TIT.1975.1055351}{IEEE Trans. Inf.
Theory {\bf IT-21}, 125 (1975).}

	\bibitem{HelstromBook} C. W. Helstrom, \textit{Quantum Detection and Estimation Theory} (Academic, New York, 1976).

	\bibitem{Ivanovic87} I. D. Ivanovic, \href{http://dx.doi.org/10.1016/0375-9601(87)90222-2}{Phys. Lett. A~{\bf 123}, 257 (1987).}

	\bibitem{Chefles98} A. Chefles and S. M. Barnett, \href{http://dx.doi.org/10.1016/S0375-9601(98)00827-5}{Phys. Lett. A {\bf 250}, 223 (1998).}

	\bibitem{Croke06} S. Croke, E. Andersson, S. M. Barnett, C. R. Gilson and J. Jeffers, \href{http://dx.doi.org/10.1103/PhysRevLett.96.070401}{\prl~\textbf{96}, 070401 (2006).}

	\bibitem{Jimenez11} O. Jim\'enez, M. A. Sol\'is-Prosser, A. Delgado, and L. Neves, \href{http://dx.doi.org/10.1103/PhysRevA.84.062315}{Phys. Rev. A {\bf 84}, 062315 (2011).}

	\bibitem{Sugimoto12} H. Sugimoto, Y. Taninaka, and A. Hayashi, \href{http://dx.doi.org/10.1103/PhysRevA.86.042311}{\pra\ {\bf 86}, 042311 (2012).}

	\bibitem{Herzog12} U. Herzog, \href{http://dx.doi.org/10.1103/PhysRevA.86.032314}{\pra~{\bf 86}, 032314 (2012).}

   \bibitem{Bagan12} E. Bagan, R. Mu\~noz-Tapia, G. A. Olivares-Renter\'ia, and J. A. Bergou, \href{http://dx.doi.org/10.1103/PhysRevA.86.040303}{\pra~{\bf 86}, 040303(R) (2012).}
	
	\bibitem{Nakahira12} K. Nakahira, T. S. Usuda, and K. Kato, \href{http://dx.doi.org/10.1103/PhysRevA.86.032316}{Phys. Rev. A {\bf 86}, 032316 (2012).}

	\bibitem{Tan08} S.-H. Tan, B. I. Erkmen, V. Giovannetti, S. Guha, S. Lloyd, L. Maccone, S. Pirandola, and J. H. Shapiro, \href{http://journals.aps.org/prl/abstract/10.1103/PhysRevLett.101.253601}{Phys. Rev. Lett. {\bf 101}, 253601 (2008)}.

	\bibitem{Pirandola11} S. Pirandola, \href{http://journals.aps.org/prl/abstract/10.1103/PhysRevLett.106.090504}{Phys. Rev. Lett. {\bf 106}, 090504 (2011)}.

	\bibitem{Nair11} R. Nair and B. J. Yen, \href{http://journals.aps.org/prl/abstract/10.1103/PhysRevLett.107.193602}{Phys. Rev. Lett. {\bf 107}, 193602 (2011)}.

	\bibitem{Lloyd11} S. Lloyd, V. Giovannetti, and L. Maccone, \href{http://journals.aps.org/prl/abstract/10.1103/PhysRevLett.106.250501}{Phys. Rev. Lett. {\bf 106}, 250501 (2011)}.

	\bibitem{Loock06} P. van Loock, T. D. Ladd, K. Sanaka, F. Yamaguchi, K. Nemoto, W. J. Munro, and Y. Yamamoto, \href{http://dx.doi.org/10.1103/PhysRevLett.96.240501}{Phys. Rev. Lett. {\bf 96}, 240501 (2006).}

	\bibitem{Ban97} M. Ban, K. Kurokawa, R. Momose, and O. Hirota, \href{http://dx.doi.org/10.1007/BF02435921}{Int. J. Theor. Phys. {\bf 36}, 1269 (1997).}

	\bibitem{Phoenix00} S. J. D. Phoenix, S. M. Barnett, and A. Chefles,  \href{http://dx.doi.org/10.1080/09500340008244056}{J. Mod. Opt. {\bf 47}, 507 (2000).}

	\bibitem{Neves12} L. Roa, A. Delgado, and I. Fuentes-Guridi, \href{http://dx.doi.org/10.1103/PhysRevA.68.022310}{\pra~{\bf 68}, 022310 (2003)}; L. Neves, M. A. Sol\'is-Prosser, A. Delgado, and O. Jim\'enez, \href{http://dx.doi.org/10.1103/PhysRevA.85.062322}{Phys. Rev. A {\bf 85}, 062322 (2012).}

	\bibitem{Prosser14} A. Delgado, L. Roa, J. C. Retamal, and C. Saavedra, \href{http://dx.doi.org/10.1103/PhysRevA.71.012303}{\pra~{\bf 71}, 012303 (2005)}; M. A. Sol\'is-Prosser, A. Delgado, O. Jim\'enez, and L. Neves, \href{http://dx.doi.org/10.1103/PhysRevA.89.012337}{Phys. Rev. A {\bf 89}, 012337 (2014).}

	\bibitem{Pati05} A. K. Pati, P. Parashar, and P. Agrawal,
\href{http://dx.doi.org/10.1103/PhysRevA.72.012329}{Phys. Rev. A {\bf 72}, 012329 (2005).}

	\bibitem{Cook07} R. L. Cook, P. J. Martin, and J. M. Geremia, \href{http://www.nature.com/nature/journal/v446/n7137/full/nature05655.html}{Nature (London) {\bf 446}, 774 (2007).}

	\bibitem{Barnett97} S. M. Barnett and E. Riis, \href{http://dx.doi.org/10.1080/09500349708230718}{J. Mod. Opt. {\bf 44}, 1061 (1997).}

	\bibitem{Clarke01} R. B. M. Clarke, V. M. Kendon, A. Chefles, S. M. Barnett, E. Riis, and M. Sasaki, \href{http://journals.aps.org/pra/abstract/10.1103/PhysRevA.64.012303}{Phys. Rev. A. {\bf 64}, 012303 (2001).}

	\bibitem{Waldherr12} G. Waldherr, A. C. Dada, P. Neumann, F. Jelezko, E. Andersson, and J. Wrachtrup, \href{http://dx.doi.org/10.1103/PhysRevLett.109.180501}{Phys. Rev. Lett. {\bf 109}, 180501 (2012).}

	\bibitem{Fujiwara03} M. Fujiwara, M. Takeoka, J. Mizuno, and M. Sasaki, \href{http://dx.doi.org/10.1103/PhysRevLett.90.167906}{Phys. Rev. Lett. {\bf 90}, 167906 (2003).}

	\bibitem{Cerf02} N. J. Cerf, M. Bourennane, A. Karlsson, and N. Gisin, \href{http://dx.doi.org/10.1103/PhysRevLett.88.127902}{Phys. Rev. Lett. {\bf 88}, 127902 (2002).}

	\bibitem{Sheridan10} L. Sheridan and V. Scarani, \href{http://dx.doi.org/10.1103/PhysRevA.82.030301}{Phys. Rev. A {\bf 82}, 030301(R) (2010).}

	\bibitem{Huber13} M. Huber and M. Paw\l{}owski, \href{http://dx.doi.org/10.1103/PhysRevA.88.032309}{\pra~{\bf 88}, 032309 (2013).}

	\bibitem{Kochen67} S. Kochen and E. P. Specker, \href{http://dx.doi.org/10.1007/978-94-010-1795-4_17}{J. Math. Mech. {\bf 17}, 59 (1967)}; G. Ca\~nas, M. Arias, S. Etcheverry, E. S. G\'omez, A. Cabello, G. B. Xavier, and G. Lima, \href{https://doi.org/10.1103/PhysRevLett.113.090404}{\prl {\bf 113}, 090404 (2014).}


	\bibitem{Vertesi10} T. V\'ertesi, S. Pironio, and N. Brunner, \href{http://dx.doi.org/10.1103/PhysRevLett.104.060401}{Phys. Rev. Lett. {\bf 104}, 060401 (2010).}

	\bibitem{Neves05} L. Neves, G. Lima, J. G. Aguirre G\'omez, C. H. Monken, C. Saavedra, and S. P\'adua,
\href{http://dx.doi.org/10.1103/PhysRevLett.94.100501}{Phys. Rev. Lett. {\bf 94}, 100501 (2005).}

	\bibitem{Chen07} P.-X. Chen, J. A. Bergou, S.-Y. Zhu, and G.-C. Guo, \href{http://dx.doi.org/10.1103/PhysRevA.76.060303}{\pra\ {\bf 76}, 060303(R) (2007).}

	\bibitem{Supplement} See Supplemental Material, which includes Refs.~\cite{Marquez07,Moreno04,Lizana10,Ivanovic81,James01,Lima11}, for more
information about the experimental setup and data analysis.

	\bibitem{Marquez07}  A. M\'arquez, I. Moreno, C. Iemmi, J. Campos, and M. J. Yzuel, \href{http://dx.doi.org/10.1117/1.2801480}{Opt. Eng. {\bf 46}, 114001 (2007).}
	
	\bibitem{Moreno04} I. Moreno, C. Iemmi, A. M\'arquez, J. Campos, and M. J. Yzuel, \href{http://dx.doi.org/10.1364/AO.43.006278}{App. Opt. {\bf 43}, 6278 (2004).}

	\bibitem{Lizana10} A. Lizana, A. M\'arquez, L. Lobato, Y. Rodange, I. Moreno, C. Iemmi, J. Campos,  \href{http://dx.doi.org/10.1364/OE.18.010581}{Opt. Express {\bf 18}, 10581 (2010).}

	\bibitem{Ivanovic81} I. D. Ivanovic, \href{http://dx.doi.org/10.1088/0305-4470/14/12/019}{J. Phys. A {\bf 14}, 3241 (1981).}

	\bibitem{James01} D. F. V. James, P. G. Kwiat, W. J. Munro, and A. G. White, \href{http://dx.doi.org/10.1103/PhysRevA.64.052312}{\pra~{\bf 64}, 052312 (2001).}
	
	\bibitem{Lima11} G. Lima, L. Neves, R. Guzm\'an, E. S. G\'omez, W. A. T. Nogueira, A. Delgado, A. Vargas, and C. Saavedra, \href{http://dx.doi.org/10.1364/OE.19.003542}{Opt. Express {\bf 19}, 3542 (2011).}

	\bibitem{Mohseni04} M. Mohseni, A. M. Steinberg, and J. A. Bergou, \href{http://dx.doi.org/10.1103/PhysRevLett.93.200403}{Phys. Rev. Lett. {\bf 93}, 200403 (2004).}

	\bibitem{Mosley06} P. J. Mosley, S. Croke, I. A. Walmsley, and S. M. Barnett, \href{http://dx.doi.org/10.1103/PhysRevLett.97.193601}{Phys. Rev. Lett. {\bf 97}, 193601 (2006).}

	\bibitem{Solis-Prosser13} M. A. Sol\'is-Prosser, A. Arias, J. J. M. Varga, L. Reb\'on, S. Ledesma, C. Iemmi, and L. Neves, \href{https://www.osapublishing.org/ol/abstract.cfm?uri=ol-38-22-4762}{Opt. Lett. {\bf 38}, 4762 (2013).}

	\bibitem{Varga14} J. J. M. Varga, L. Reb\'on, M. A. Sol\'is-Prosser, L. Neves, S. Ledesma, and C. Iemmi, 
\href{http://iopscience.iop.org/article/10.1088/0953-4075/47/22/225504?fromSearchPage=true}{J. Phys. B {\bf 47},  225504 (2014).}

	\bibitem{Solis-Prosser11} M. A. Sol\'{i}s-Prosser and L. Neves, 
\href{http://journals.aps.org/pra/abstract/10.1103/PhysRevA.84.012330}{Phys. Rev. A {\bf 84}, 012330 (2011).}

	\bibitem{Chefles02} A. Chefles, \href{https://doi.org/10.1103/PhysRevA.66.042325}{Phys. Rev. A {\bf 66}, 042325  (2002).}

	\bibitem{Comment} The way we have parametrized the state coefficients for $D=2$ to 9 and given that they were randomly selected for $D=10$ to 21 ensure that our experiment did not restrict itself to a particular subset of symmetric states, encompassing, instead, sets constructed from arbitrary fiducial states [$|\psi_0\rangle$ in Eq.~(\ref{eq:sym-states})].

	\bibitem{Fickler13} R. Fickler, M. Krenn, R. Lapkiewicz, S. Ramelow, and A. Zeilinger, \href{doi:10.1038/srep01914}{Sci. Rep. {\bf 3}, 1914 (2013).}
 
	\bibitem{Zhang09} L. Zhang, L. Neves, J. S. Lundeen, and I. A. Walmsley, \href{https://doi.org/10.1088/0953-4075/42/11/114011}{J. Phys. B {\bf 42}, 114011 (2009).}

	\bibitem{Scarcella13} C. Scarcella, A. Tosi, F. Villa, S. Tisa, and F. Zappa, \href{http://dx.doi.org/10.1063/1.4850677}{Rev. Sci. Instrum.~{\bf 84}, 123112 (2013).}

	\bibitem{Unter16} M. Untern\"ahrer, B. Bessire, L. Gasparini, D. Stoppa, and A. Stefanov, \href{https://doi.org/10.1364/OE.24.028829}{Opt. Express {\bf 24}, 28829 (2016).}
	
	\bibitem{CheflesDist} A. Chefles, \href{http://dx.doi.org/10.1103/PhysRevA.66.042325}{Phys. Rev. A {\bf 66}, 042325 (2002).}
	
\end{thebibliography}
